\documentclass[pra,superscriptaddress,twocolumn,floatfix,nofootinbib]{revtex4-1}
\usepackage{graphicx,color}
\usepackage{amsmath,amssymb,bm}
\usepackage{simplewick}
\usepackage{array}
\usepackage{dsfont} 
\usepackage{tikz,tikz-3dplot,pgfplots}				
\usepackage{braket}		

\newcommand{\ccaption}[2]{\caption[#1]{\textit{#1.} #2}}
\newcommand{\ii}{\mathrm{i}} 
\newcommand{\ee}{e}
\newcommand{\Mod}[1]{(\mathrm{mod}\ #1)}

\usepackage[plainpages=false,pdfpagelabels,colorlinks=true,linkcolor=red,urlcolor=blue,citecolor=blue,pdftitle={Entanglement Entropy of Typical Eigenstates of Translationally Invariant Quadratic Hamiltonians},pdfauthor={},pdfdisplaydoctitle=true,pdfduplex=DuplexFlipLongEdge]{hyperref}

\definecolor{darkred}{rgb}{0.90,0.2,0.2}
\definecolor{darkgreen}{rgb}{0,0.60,.2}
\definecolor{darkblue}{rgb}{0.1,0.3,1}
\definecolor{grey}{cmyk}{0,0,0,0.25}
\definecolor{orange}{cmyk}{0,0.6,0.8,0}

\newcommand{\conf}[4]{
	\vcenter{\hbox{
			\begin{tikzpicture}
			\draw[white] (-.3,-.4) rectangle (.3,.4);
			\pgfmathsetmacro{\r}{.3}%
			\coordinate (c) at (-45:\r);
			\coordinate (b) at (45:\r);
			\coordinate (a) at (135:\r);
			\coordinate (d) at (-135:\r);
			\coordinate (o) at (0,0);
			\draw (o) circle (\r);
			\draw[black,very thick] (#1) .. controls ($1/4*(#1)+1/4*(#2)$) and ($1/4*(#1)+1/4*(#2)$) .. (#2);
			\draw[white] (#1) .. controls ($1/4*(#1)+1/4*(#2)$) and ($1/4*(#1)+1/4*(#2)$) .. (#2);
			\draw[black,very thick] (#2) .. controls ($1/4*(#2)+1/4*(#3)$) and ($1/4*(#2)+1/4*(#3)$) .. (#3);
			\draw[white] (#2) .. controls ($1/4*(#2)+1/4*(#3)$) and ($1/4*(#2)+1/4*(#3)$) .. (#3);
			\draw[black,very thick] (#3) .. controls ($1/4*(#3)+1/4*(#4)$) and ($1/4*(#3)+1/4*(#4)$) .. (#4);
			\draw[white] (#3) .. controls ($1/4*(#3)+1/4*(#4)$) and ($1/4*(#3)+1/4*(#4)$) .. (#4);
			\draw[black,very thick] (#4) .. controls ($1/4*(#4)+1/4*(#1)$) and ($1/4*(#4)+1/4*(#1)$) .. (#1);
			\draw[white] (#4) .. controls ($1/4*(#4)+1/4*(#1)$) and ($1/4*(#4)+1/4*(#1)$) .. (#1);
			\fill (a) circle[radius=1pt];
			\fill (b) circle[radius=1pt];
			\fill (c) circle[radius=1pt];
			\fill (d) circle[radius=1pt];
			\end{tikzpicture}	
	}}
}
\newcommand{\contt}[4]{
	\vcenter{\hbox{
			\begin{tikzpicture}
			\draw[white] (-.3,-.4) rectangle (.3,.4);
			\pgfmathsetmacro{\r}{.3}%
			\coordinate (c) at (-45:\r);
			\coordinate (b) at (45:\r);
			\coordinate (a) at (135:\r);
			\coordinate (d) at (-135:\r);
			\coordinate (o) at (0,0);
			\draw (o) circle (\r);
			\draw[black,very thick] (#1) .. controls ($1/4*(#1)+1/4*(#2)$) and ($1/4*(#1)+1/4*(#2)$) .. (#2);
			\draw[white] (#1) .. controls ($1/4*(#1)+1/4*(#2)$) and ($1/4*(#1)+1/4*(#2)$) .. (#2);
			\draw[black,very thick] (#3) .. controls ($1/4*(#3)+1/4*(#4)$) and ($1/4*(#3)+1/4*(#4)$) .. (#4);
			\draw[white] (#3) .. controls ($1/4*(#3)+1/4*(#4)$) and ($1/4*(#3)+1/4*(#4)$) .. (#4);
			\fill (a) circle[radius=1pt];
			\fill (b) circle[radius=1pt];
			\fill (c) circle[radius=1pt];
			\fill (d) circle[radius=1pt];
			\end{tikzpicture}	
	}}
}
\newcommand{\conttt}[6]{
	\vcenter{\hbox{
			\begin{tikzpicture}
			\draw[white] (-.3,-.4) rectangle (.3,.4);
			\pgfmathsetmacro{\r}{.3}%
			\coordinate (a) at (90:\r);
			\coordinate (b) at (30:\r);
			\coordinate (c) at (-30:\r);
			\coordinate (d) at (-90:\r);
			\coordinate (e) at (-150:\r);
			\coordinate (f) at (-210:\r);
			\coordinate (o) at (0,0);
			\draw (o) circle (\r);
			\draw[black,very thick] (#1) .. controls ($1/4*(#1)+1/4*(#2)$) and ($1/4*(#1)+1/4*(#2)$) .. (#2);
			\draw[white] (#1) .. controls ($1/4*(#1)+1/4*(#2)$) and ($1/4*(#1)+1/4*(#2)$) .. (#2);
			\draw[black,very thick] (#3) .. controls ($1/4*(#3)+1/4*(#4)$) and ($1/4*(#3)+1/4*(#4)$) .. (#4);
			\draw[white] (#3) .. controls ($1/4*(#3)+1/4*(#4)$) and ($1/4*(#3)+1/4*(#4)$) .. (#4);
			\draw[black,very thick] (#5) .. controls ($1/4*(#5)+1/4*(#6)$) and ($1/4*(#5)+1/4*(#6)$) .. (#6);
			\draw[white] (#5) .. controls ($1/4*(#5)+1/4*(#6)$) and ($1/4*(#5)+1/4*(#6)$) .. (#6);
			\fill (a) circle[radius=1pt];
			\fill (b) circle[radius=1pt];
			\fill (c) circle[radius=1pt];
			\fill (d) circle[radius=1pt];
			\fill (e) circle[radius=1pt];
			\fill (f) circle[radius=1pt];
			\end{tikzpicture}	
	}}
}
\newcommand{\contttt}[8]{
	\vcenter{\hbox{
			\begin{tikzpicture}
			\draw[white] (-.3,-.4) rectangle (.3,.4);
			\pgfmathsetmacro{\r}{.3}%
			\coordinate (c) at (22.5:\r);
			\coordinate (b) at (67.5:\r);
			\coordinate (a) at (112.5:\r);
			\coordinate (h) at (157.5:\r);
			\coordinate (g) at (202.5:\r);
			\coordinate (f) at (247.5:\r);
			\coordinate (e) at (292.5:\r);
			\coordinate (d) at (337.5:\r);
			\coordinate (o) at (0,0);
			\draw (o) circle (\r);
			\draw[black,very thick] (#1) .. controls ($1/4*(#1)+1/4*(#2)$) and ($1/4*(#1)+1/4*(#2)$) .. (#2);
			\draw[white] (#1) .. controls ($1/4*(#1)+1/4*(#2)$) and ($1/4*(#1)+1/4*(#2)$) .. (#2);
			\draw[black,very thick] (#3) .. controls ($1/4*(#3)+1/4*(#4)$) and ($1/4*(#3)+1/4*(#4)$) .. (#4);
			\draw[white] (#3) .. controls ($1/4*(#3)+1/4*(#4)$) and ($1/4*(#3)+1/4*(#4)$) .. (#4);
			\draw[black,very thick] (#5) .. controls ($1/4*(#5)+1/4*(#6)$) and ($1/4*(#5)+1/4*(#6)$) .. (#6);
			\draw[white] (#5) .. controls ($1/4*(#5)+1/4*(#6)$) and ($1/4*(#5)+1/4*(#6)$) .. (#6);
			\draw[black,very thick] (#7) .. controls ($1/4*(#7)+1/4*(#8)$) and ($1/4*(#7)+1/4*(#8)$) .. (#8);
			\draw[white] (#7) .. controls ($1/4*(#7)+1/4*(#8)$) and ($1/4*(#7)+1/4*(#8)$) .. (#8);
			\fill (a) circle[radius=1pt];
			\fill (b) circle[radius=1pt];
			\fill (c) circle[radius=1pt];
			\fill (d) circle[radius=1pt];
			\fill (e) circle[radius=1pt];
			\fill (f) circle[radius=1pt];
			\fill (g) circle[radius=1pt];
			\fill (h) circle[radius=1pt];
			\end{tikzpicture}	
	}}
}
\newcommand{\conff}[8]{
	\vcenter{\hbox{
			\begin{tikzpicture}
			\draw[white] (-.3,-.4) rectangle (.3,.4);
			\pgfmathsetmacro{\r}{.3}%
			\coordinate (c) at (22.5:\r);
			\coordinate (b) at (67.5:\r);
			\coordinate (a) at (112.5:\r);
			\coordinate (h) at (157.5:\r);
			\coordinate (g) at (202.5:\r);
			\coordinate (f) at (247.5:\r);
			\coordinate (e) at (292.5:\r);
			\coordinate (d) at (337.5:\r);
			\coordinate (o) at (0,0);
			\draw (o) circle (\r);
			\draw[black,very thick] (#1) .. controls ($1/4*(#1)+1/4*(#2)$) and ($1/4*(#1)+1/4*(#2)$) .. (#2);
			\draw[white] (#1) .. controls ($1/4*(#1)+1/4*(#2)$) and ($1/4*(#1)+1/4*(#2)$) .. (#2);
			\draw[black,very thick] (#2) .. controls ($1/4*(#2)+1/4*(#3)$) and ($1/4*(#2)+1/4*(#3)$) .. (#3);
			\draw[white] (#2) .. controls ($1/4*(#2)+1/4*(#3)$) and ($1/4*(#2)+1/4*(#3)$) .. (#3);
			\draw[black,very thick] (#3) .. controls ($1/4*(#3)+1/4*(#4)$) and ($1/4*(#3)+1/4*(#4)$) .. (#4);
			\draw[white] (#3) .. controls ($1/4*(#3)+1/4*(#4)$) and ($1/4*(#3)+1/4*(#4)$) .. (#4);
			\draw[black,very thick] (#4) .. controls ($1/4*(#4)+1/4*(#1)$) and ($1/4*(#4)+1/4*(#1)$) .. (#1);
			\draw[white] (#4) .. controls ($1/4*(#4)+1/4*(#1)$) and ($1/4*(#4)+1/4*(#1)$) .. (#1);
			\draw[gray,very thick] (#5) .. controls ($1/4*(#5)+1/4*(#6)$) and ($1/4*(#5)+1/4*(#6)$) .. (#6);
			\draw[white] (#5) .. controls ($1/4*(#5)+1/4*(#6)$) and ($1/4*(#5)+1/4*(#6)$) .. (#6);
			\draw[gray,very thick] (#6) .. controls ($1/4*(#6)+1/4*(#7)$) and ($1/4*(#6)+1/4*(#7)$) .. (#7);
			\draw[white] (#6) .. controls ($1/4*(#6)+1/4*(#7)$) and ($1/4*(#6)+1/4*(#7)$) .. (#7);
			\draw[gray,very thick] (#7) .. controls ($1/4*(#7)+1/4*(#8)$) and ($1/4*(#7)+1/4*(#8)$) .. (#8);
			\draw[white] (#7) .. controls ($1/4*(#7)+1/4*(#8)$) and ($1/4*(#7)+1/4*(#8)$) .. (#8);
			\draw[gray,very thick] (#8) .. controls ($1/4*(#8)+1/4*(#5)$) and ($1/4*(#8)+1/4*(#5)$) .. (#5);
			\draw[white] (#8) .. controls ($1/4*(#8)+1/4*(#5)$) and ($1/4*(#8)+1/4*(#5)$) .. (#5);
			\fill (a) circle[radius=1pt];
			\fill (b) circle[radius=1pt];
			\fill (c) circle[radius=1pt];
			\fill (d) circle[radius=1pt];
			\fill (e) circle[radius=1pt];
			\fill (f) circle[radius=1pt];
			\fill (g) circle[radius=1pt];
			\fill (h) circle[radius=1pt];
			\end{tikzpicture}	
	}}
}
\newcommand{\conttf}[8]{
	\vcenter{\hbox{
			\begin{tikzpicture}
			\draw[white] (-.3,-.4) rectangle (.3,.4);
			\pgfmathsetmacro{\r}{.3}%
			\coordinate (c) at (22.5:\r);
			\coordinate (b) at (67.5:\r);
			\coordinate (a) at (112.5:\r);
			\coordinate (h) at (157.5:\r);
			\coordinate (g) at (202.5:\r);
			\coordinate (f) at (247.5:\r);
			\coordinate (e) at (292.5:\r);
			\coordinate (d) at (337.5:\r);
			\coordinate (o) at (0,0);
			\draw (o) circle (\r);
			\draw[black,very thick] (#1) .. controls ($1/4*(#1)+1/4*(#2)$) and ($1/4*(#1)+1/4*(#2)$) .. (#2);
			\draw[white] (#1) .. controls ($1/4*(#1)+1/4*(#2)$) and ($1/4*(#1)+1/4*(#2)$) .. (#2);
			\draw[black,very thick] (#4) .. controls ($1/4*(#4)+1/4*(#3)$) and ($1/4*(#4)+1/4*(#3)$) .. (#3);
			\draw[white] (#4) .. controls ($1/4*(#4)+1/4*(#3)$) and ($1/4*(#4)+1/4*(#3)$) .. (#3);
			\draw[gray,very thick] (#5) .. controls ($1/4*(#5)+1/4*(#6)$) and ($1/4*(#5)+1/4*(#6)$) .. (#6);
			\draw[white] (#5) .. controls ($1/4*(#5)+1/4*(#6)$) and ($1/4*(#5)+1/4*(#6)$) .. (#6);
			\draw[gray,very thick] (#6) .. controls ($1/4*(#6)+1/4*(#7)$) and ($1/4*(#6)+1/4*(#7)$) .. (#7);
			\draw[white] (#6) .. controls ($1/4*(#6)+1/4*(#7)$) and ($1/4*(#6)+1/4*(#7)$) .. (#7);
			\draw[gray,very thick] (#7) .. controls ($1/4*(#7)+1/4*(#8)$) and ($1/4*(#7)+1/4*(#8)$) .. (#8);
			\draw[white] (#7) .. controls ($1/4*(#7)+1/4*(#8)$) and ($1/4*(#7)+1/4*(#8)$) .. (#8);
			\draw[gray,very thick] (#8) .. controls ($1/4*(#8)+1/4*(#5)$) and ($1/4*(#8)+1/4*(#5)$) .. (#5);
			\draw[white] (#8) .. controls ($1/4*(#8)+1/4*(#5)$) and ($1/4*(#8)+1/4*(#5)$) .. (#5);
			\fill (a) circle[radius=1pt];
			\fill (b) circle[radius=1pt];
			\fill (c) circle[radius=1pt];
			\fill (d) circle[radius=1pt];
			\fill (e) circle[radius=1pt];
			\fill (f) circle[radius=1pt];
			\fill (g) circle[radius=1pt];
			\fill (h) circle[radius=1pt];
			\end{tikzpicture}
	}}
}

\begin{document}

\title{Average eigenstate entanglement entropy of the XY chain in a transverse field\\and its universality for translationally invariant quadratic fermionic models}

\author{Lucas Hackl}
\affiliation{Max  Planck  Institute of Quantum Optics, Hans-Kopfermann-Stra{\ss}e 1, D-85748 Garching bei M\"unchen, Germany}
\affiliation{Institute for Gravitation and the Cosmos, The Pennsylvania State University, University Park, PA 16802, USA}
\affiliation{Department of Physics, The Pennsylvania State University, University Park, PA 16802, USA}
\author{Lev Vidmar}
\affiliation{Department of Theoretical Physics, J. Stefan Institute, SI-1000 Ljubljana, Slovenia}
\author{Marcos Rigol}
\affiliation{Department of Physics, The Pennsylvania State University, University Park, PA 16802, USA}
\author{Eugenio Bianchi}
\affiliation{Institute for Gravitation and the Cosmos, The Pennsylvania State University, University Park, PA 16802, USA}
\affiliation{Department of Physics, The Pennsylvania State University, University Park, PA 16802, USA}

\begin{abstract}
We recently showed [\href{https://journals.aps.org/prl/abstract/10.1103/PhysRevLett.121.220602}{Phys. Rev. Lett.~{\bf 121}, 220602 (2018)}] that the average bipartite entanglement entropy of all energy eigenstates of the quantum Ising chain exhibits a universal (for translationally invariant quadratic fermionic models) leading term that scales linearly with the subsystem's volume, while in the thermodynamic limit the first subleading correction does not vanish at the critical field (it only depends on the ratio $f$ between the volume of the subsystem and volume of the system) and vanishes otherwise. Here we show, analytically for bounds and numerically for averages, that the same remains true for the spin-1/2 XY chain in a transverse magnetic field. We then tighten the bounds for the coefficient of the universal volume-law term, which is a concave function of $f$. We develop a systematic approach to compute upper and lower bounds, and provide explicit analytic expressions for up to the fourth order bounds.
\end{abstract}

\maketitle


\section{Introduction}

Entanglement is a hallmark of quantum theory. The correlations between two entangled quantum systems that are in an overall pure state cannot be explained by a local classical theory~\cite{epr_35, bell2001einstein}. Different measures of entanglement have been extensively studied in the context of black hole physics~\cite{Sorkin:2014kta, Bombelli:1986rw, srednicki_93, holzhey_larsen_94}, holography~\cite{Ryu:2006bv, Lewkowycz:2013nqa}, ground states of condensed matter systems~\cite{white_92, schollwoeck_05, schollwoeck_11}, and quantum information science~\cite{nielsen_chuang_10}. Recently, the first measurements of an entanglement quantifier were done with ultracold atoms in an optical lattice~\cite{islam_ma_15, kaufman_tai_16}.

Given a quantum system with Hilbert space $\mathcal{H}=\mathcal{H}_A\otimes \mathcal{H}_B$, which can be decomposed as a tensor product of $\mathcal{H}_A$ and $\mathcal{H}_B$, the bipartite von Neumann entanglement entropy (or, in short, the entanglement entropy) of a pure state $\ket{\psi}\in\mathcal{H}$ is defined as
\begin{equation} \label{def_SvN}
 S_A(\ket{\psi})=-\mathrm{Tr}(\hat\rho_A\ln\hat\rho_A),\ \ \text{with}\ \ \hat\rho_A=\mathrm{Tr}_{\mathcal{H}_B}\!\ket{\psi}\!\bra{\psi},
\end{equation}
where a partial trace is taken over $\mathcal{H}_B$ to find the reduced (mixed) state $\hat\rho_A$ of $\ket{\psi}$. Here, we are interested in universal features of the entanglement entropy of eigenstates of lattice Hamiltonians for a bipartition into two blocks, in which $L_A$ is the volume (number of lattice sites) of the smaller block. In particular, we investigate --  both analytically and numerically -- the average eigenstate entanglement entropy (over all eigenstates) of the spin-1/2 XY chain in a transverse magnetic field~\cite{lieb61}, at and away the critical field for which the quantum phase transition occurs in the ground state.

The ground-state entanglement entropy of local quadratic Hamiltonians has been extensively studied in recent years~\cite{amico_fazio_08, peschel_eisler_09, calabrese_cardy_09, eisert_cramer_10, audenaert_eisert_02, osterloh_amico_2002, osborne_nielsen_02, vidal_latorre_03, latorre_rico_04, calabrese_cardy_04, jin_korepin_04, calabrese_cardy_04, peschel_04, its_jin_05, wolf_06, gioev_klich_06, barthel_chung_06, li_ding_06, cramer_eisert_07, franchini_its_07, hastings_07, ding_brayali_08}, and several remarkable analytic results have been obtained. In contrast, despite growing interest~\cite{alba09, moelter_barthel_14, storms14, beugeling15, lai15, nandy_sen_16, vidmar_hackl_17, riddell_mueller_18, zhang_vidmar_18, vidmar2018volume, murciano_ruggiero_18}, an equivalent analytic understanding of the behavior of the entanglement entropy of excited eigenstates is still lacking. Recent analytic results include: (i) bounds for the average entanglement entropy over all eigenstates of translationally invariant quadratic fermionic models~\cite{vidmar_hackl_17}, and (ii) the average entanglement entropy over eigenstates of the quadratic part of the Sachdev-Ye-Kitaev model~\cite{liu_chen_18}. Those results have provided proofs that the leading term of the average entanglement entropy $\langle S_A\rangle$ in the corresponding quadratic models scales with the volume, $\langle S_A\rangle \propto L_A$. In another recent study~\cite{vidmar2018volume}, it was shown that for the quantum Ising chain the leading term of $\langle S_A\rangle$ is independent of the Hamiltonian parameters, and it was conjectured to be universal for all translationally invariant quadratic fermionic Hamiltonians. Furthermore, it was shown that the subleading term of $\langle S_A\rangle$ in the quantum Ising chain depends on whether the system is at the critical field or not. Generalizing these results to the spin-1/2 XY chain in a transverse field is the first goal of this paper.

The bounds in Ref.~\cite{vidmar_hackl_17}, as well as numerical results for noninteracting fermions~\cite{storms14, vidmar_hackl_17}, have shown that the coefficient of the volume-law of the average eigenstate entanglement entropy of translationally invariant quadratic fermionic Hamiltonians is smaller than the theoretical maximum. The latter is attained by the (Haar measure) average over pure states in the Hilbert space~\cite{page93} and by the average over eigenstates of quantum-chaotic Hamiltonians~\cite{vidmar_rigol_17}. Motivated by questions about thermalization in generic isolated quantum systems~\cite{dalessio_kafri_16}, the entanglement properties of excited eigenstates of quantum-chaotic Hamiltonians have been the focus of much recent work~\cite{deutsch_10, santos_polkovnikov_12, hamma_santra_12, deutsch13, alba15, beugeling15, yang_chamon_15, vivo_pato_16, vidmar_rigol_17, garrison_grover_18, dymarsky_lashkari_18, nakagawa_watanabe_18, huang_19, luitz2016long, lu2017renyi}. In the context of thermalization, it is of particular interest to understand how entanglement properties differ between many-body energy eigenstates of quantum chaotic systems (systems that are expected to thermalize under unitary dynamics) and many-body energy eigenstates of integrable systems (systems that are not expected to thermalize under unitary dynamics).

An important property of the volume-law coefficient of the average eigenstate entanglement entropy in translationally invariant quadratic fermionic Hamiltonians (which are a class of particularly simple integrable models) is that it is a concave function of the ratio between the volume of the subsystem $L_A$ and the volume of the system $L$ (in short, the subsystem fraction $f=L_A/L$) for $f>0$. To learn more about that concave function, the second goal of this paper is to sharpen the bounds reported in Ref.~\cite{vidmar_hackl_17} for the volume-law coefficient.

The presentation is structured as follows. In Sec.~\ref{sec:xy-model}, we study average entanglement entropies of eigenstates of the spin-1/2 XY chain in a transverse field. Bounds are determined analytically, while exact averages are computed numerically. In Sec.~\ref{sec:xx-model}, we tighten the bounds reported in Ref.~\cite{vidmar_hackl_17} for the conjectured universal leading term of the average entanglement entropy in translationally invariant quadratic fermionic Hamiltonians. In particular, we provide explicit analytical expressions for up to the fourth order bounds in our perturbative expansion for noninteracting fermions. A summary and discussion of our results is provided in Sec.~\ref{sec:conclusion}. In Appendix~\ref{app:evenodd}, we discuss details about the effect of eigenstates in the fermionic even/odd particle-number sectors, which are not eigenstates of the spin Hamiltonian, in calculations of averages. Appendix~\ref{app1} provides a comprehensive review of our perturbative method, based on a generalized Wick's theorem, and reports details of the calculations of up to the fourth order bounds.

\begin{figure}
\includegraphics[width=\linewidth]{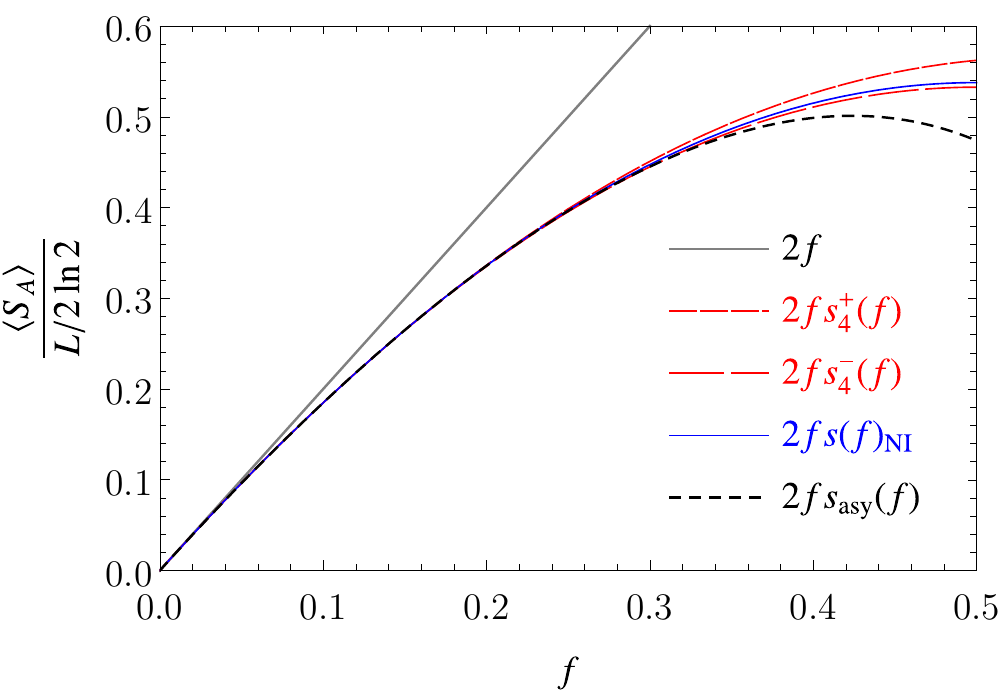}
\vspace{-7mm}
\ccaption{Average entanglement entropy as a function of the subsystem fraction}{Results are shown for the maximum  entanglement entropy $2f$, the fourth order bounds $2f s_4^\pm(f)$ from Eqs.~(\ref{def_sp4}) and~(\ref{def_s4m}), the numerical average $2f s(f)_{\rm NI}$ for noninteracting fermions in a chain with $L=36$, and the asymptotic result for the average $2f s_{\rm asy}(f)$ from Eq.~\eqref{def_sasy}.}
\label{fig:intro}
\end{figure}

Figure~\ref{fig:intro} reports the main results of Sec.~\ref{sec:xx-model}, which are among the main results of this work. In Fig.~\ref{fig:intro}, we compare the average entanglement entropy, obtained numerically, with the fourth order upper and lower bounds, and with the exact asymptotic expression for the average for up to $f^4$. That figure shows that the fourth order bounds, specially the lower one, are very tight, and that the asymptotic expression is accurate up to $f\approx 0.3$.

\section{XY model in a transverse field}\label{sec:xy-model}
We consider the spin-1/2 XY model in a transverse field in one dimension, in systems with $L$ sites and periodic boundary conditions~\cite{lieb61}. We denote sites as $j=1,\ldots, L$, with site $L+1$ identified as site $1$. For simplicity, we assume that $L$ is an even integer. The Hamiltonian can be written as
\begin{equation}\label{eq:HXY1}
 \hat{H}_{\mathrm{XY}}=-J\sum^L_{j =1}\left[(1+\gamma)\hat{S}^{\mathrm{X}}_j \hat{S}^{\mathrm{X}}_{j +1}+(1-\gamma)\hat{S}^{\mathrm{Y}}_j \hat{S}^{\mathrm{Y}}_{j +1}\right]-h\hat{S}^{\mathrm{Z}}_j,
\end{equation}
where $\hat{S}^{\mathrm{X}}_j$, $\hat{S}^{\mathrm{Y}}_j$, and $\hat{S}^{\mathrm{Z}}_j$ are standard spin-1/2 operators, $J$ is the interaction strength, $h$ is the strength of the transverse field, and $0\leq\gamma\leq 1$ is the anisotropy parameter.

This model can be solved by applying first a Jordan-Wigner transformation and subsequently a Bogoliubov transformation. The resulting Hamiltonian can be written as sum $\hat{H}_{\mathrm{XY}}=\hat{H}_{\mathrm{XY}}^{+}\mathcal{P}^++\hat{H}_{\mathrm{XY}}^{-}\mathcal{P}^-$ where $\mathcal{P}^\pm$ are projectors onto Hilbert sub spaces $\mathcal{H}^\pm$ with even ($+$) and odd ($-$) number of excitations. The individual terms
	\begin{align}
		\hat{H}^\pm_{\mathrm{XY}}=\sum_{\kappa\in\mathcal{K}^\pm}\varepsilon_\kappa\ \left(\hat{\eta}_\kappa^\dagger\hat{\eta}_\kappa-\frac{1}{2}\right)\,.\label{eq:HXYn}
	\end{align}
are quadratic in fermionic creation and annihilation operators $\hat{\eta}_k^\dagger$ and $\hat{\eta}_\kappa^\dagger$ with
\begin{align}
\kappa\in\mathcal{K}^+&=\left\{\frac{\pi}{L}+\frac{2\pi k}{L}\;\Bigg|\;k\in\mathbb{Z}\,,-\frac{L}{2}\leq k<\frac{L}{2}\right\}\,, \label{def_Kplus}\\
\mathcal{K}^-&=\left\{\frac{2\pi k}{L}\;\Bigg|\;k\in\mathbb{Z}\,,-\frac{L}{2}< k\leq \frac{L}{2}\right\}\,. \label{def_Kminus}
\end{align}
We can express them in terms of localized creation and annihilation operators $\hat{f}_j^\dagger$ and $\hat{f}_j^\dagger$ at site $j$ via
\begin{align}
	\hat\eta_\kappa=\frac{1}{\sqrt{L}}\sum^L_{j=1}\left(u_\kappa\, e^{\ii \kappa j}\hat{f}_j-v^*_\kappa \,e^{-\ii \kappa j}\hat{f}^\dagger_j\,\right)\,,
\end{align}
where the Bogoliubov coefficients and energies are
\begin{align}
\begin{split}
u_\kappa&=\frac{\varepsilon_\kappa+a_\kappa}{\sqrt{2\varepsilon_\kappa(\varepsilon_\kappa+a_\kappa)}}\,,\quad v_\kappa=\frac{\ii b_\kappa}{\sqrt{2\varepsilon_\kappa(\varepsilon_\kappa+a_\kappa)}} \,,\\
\varepsilon_\kappa&=\sqrt{h^2+2h J\cos(\kappa)+J^2+(\gamma^2-1)J^2\sin(\kappa)^2}\,.	\hspace{-2em}{}
\end{split}
\end{align}
A detailed derivation of this result can be found in Appendix~\ref{app:XYmodel}.

We will compare the results for the spin-1/2 XY model in a transverse field with those for noninteracting fermions, whose Hamiltonian has the form
\begin{equation} \label{def_Hfree}
 \hat{H}_{\mathrm{NI}} = - \sum^L_{j =1}(\hat{f}_j ^\dagger\hat{f}_{j+1}+\text{H.c.}) \, .
\end{equation}
This is a special case of the more general class of quadratic fermionic Hamiltonians with translational invariance (see Eq.~\ref{eq:Hquadratic}).

\subsection{Eigenstate entanglement entropy}

The eigenstates $\ket{\{n_\kappa\}}$ are fermionic Gaussian states~\cite{peschel_03, peschel_eisler_09, dierckx_fannes_08, hackl_bianchi_17}. For simplicity, we denote them as $|\psi\rangle$ in what follows. We compute the entanglement entropy $S_A(\ket{\psi})$, see Eq.~(\ref{def_SvN}), with respect to the Hilbert space decomposition $\mathcal{H} = \mathcal{H}_A\otimes\mathcal{H}_B$ by splitting $L$ sites into a connected piece of length $L_A$ and its complement with $L_B=L-L_A$ sites.

An efficient method of computing $S_A(\ket{\psi})$ is based on parametrizing Gaussian states in terms of the linear complex structure $\mathbb{J}(|\psi\rangle)$, defined as
\begin{equation} \label{def_J}
 \ii \mathbb{J}(|\psi\rangle) \equiv\left(\begin{array}{c|c} \langle\psi|\hat{f}_j^\dagger \hat{f}_l-\hat{f}_l\hat{f}_j ^\dagger|\psi\rangle & \langle\psi|\hat{f}_j ^\dagger \hat{f}_l^\dagger-\hat{f}_l^\dagger \hat{f}_j ^\dagger|\psi\rangle\\ [1mm] \hline\\ [-2.75mm]
 \langle\psi|\hat{f}_l \hat{f}_j -\hat{f}_j  \hat{f}_l|\psi\rangle & \langle\psi|\hat{f}_j  \hat{f}_l^\dagger-\hat{f}_l^\dagger\hat{f}_j |\psi\rangle
\end{array}\right)\,,
\end{equation}
where $j$ and $l$ run over all sites on the system, so that each matrix entry in Eq.~(\ref{def_J}) represents a block of size $L \times L$, i.e., $\ii \mathbb{J}(|\psi\rangle)$ is a $2L \times 2L$ matrix. For simplicity, in what follows we drop the state label $|\psi\rangle$ in $\ii \mathbb{J}$. The entanglement entropy of $\ket{\psi}$, associated to the subsystem $\mathcal{H}_A$, can be computed using the formula
\begin{equation}
 S_A(\ket{\psi}) = - \mathrm{Tr}\left[ \left(\frac{\mathds{1}_A+[\ii \mathbb{J}]_A}{2}\right)\ln \left(\frac{\mathds{1}_A+[\ii \mathbb{J}]_A}{2}\right) \right], \label{def_Spsi} 
\end{equation}
where $\mathds{1}_A$ and $[\ii \mathbb{J}]_A$ are restrictions to the $2L_A \times 2L_A$ matrices with rows and columns associated to $j$ and $l$ in the subsystem only. The advantage of using this representation to evaluate Eq.~(\ref{def_SvN}) for Gaussian states relies on the reduction from working with $2^{L_A} \times 2^{L_A}$ many-body matrices to working with $2L_A \times 2L_A$ one-body matrices. 

In Ref.~\cite{vidmar_hackl_17}, we took a step forward in simplifying the evaluation of Eq.~(\ref{def_Spsi}). It is based on the fact that the eigenvalues of $[\ii \mathbb{J}]_A$ are limited to the interval $[-1,1]$. As a result, $S_A(\ket{\psi})$ can be expressed as a converging series
\begin{equation}
 S_A(\ket{\psi}) =L_A\ln{2}-\sum^\infty_{n=1}\frac{\mathrm{Tr}[\ii \mathbb{J}]_A^{2n}}{4n(2n-1)} \,. \label{def_Sseries}
\end{equation}
Since the traces of even powers of $[\ii \mathbb{J}]_A$ are all positive, one can modify the series in Eq.~(\ref{def_Sseries}) at any order $m$ to find upper and lower bounds $S^{m+}_A(\ket{\psi})$ and $S^{m-}_A(\ket{\psi})$, respectively, given by
\begin{align} \label{def_Splus}
 S^{m+}_A(\ket{\psi})&=L_A\ln{2}-\sum^m_{n=1}\frac{\mathrm{Tr}[\ii \mathbb{J}]_A^{2n}}{4n(2n-1)},\\
 S^{m-}_A(\ket{\psi})&=L_A\ln{2}-\sum^m_{n=1}\frac{\mathrm{Tr}[\ii \mathbb{J}]_A^{2n}}{4n(2n-1)}-\hspace{-2mm}\sum^\infty_{n=m+1}\frac{\mathrm{Tr}[\ii \mathbb{J}]_A^{2m}}{4n(2n-1)}.\label{def_Sminus}
\end{align}

We are interested in the average $\langle S_A\rangle$ over all eigenstates $\ket{\{n_\kappa\}}$ in the thermodynamic limit ($L\to\infty$), with the average defined as
\begin{equation} \label{def_Savr}
 \langle S_A\rangle=\frac{1\,}{2^{L}}\sum_{\{n_\kappa\}} S_A\big(\ket{\{n_\kappa\}}\big)\,,
\end{equation}
namely, one first computes the entanglement entropy $S_A(\ket{\{n_\kappa\}})$ of each eigenstate $\ket{\{n_\kappa\}}$, and then takes the average over all eigenstates. If the Hamiltonian has a non-degenerate spectrum, then the sum over the numbers $\{n_\kappa\}$ coincides with the sum over eigenstates of energy $E=\sum_{\kappa\in\mathcal{K}^\pm}n_\kappa\varepsilon_\kappa$. If, on the other hand, the Hamiltonian has a degenerate spectrum, the sum over $\{n_\kappa\}$ resolves the degeneracy within each eigenspace by introducing the orthonormal basis of Gaussian states $\ket{\{n_\kappa\}}$. This procedure is equivalent to removing the degeneracy by a vanishingly small perturbation of the $\varepsilon_\kappa$'s, such that no two states $\ket{\{n_\kappa\}}$ have the same energy.

In Ref.~\cite{vidmar_hackl_17}, we proved that the leading term of $\langle S_A\rangle$ for translationally invariant quadratic fermionic models exhibits a volume-law scaling. We also proved that, in the thermodynamic limit when $f\rightarrow0$, where $f=L_A/L$, the overwhelming majority of eigenstates exhibit the maximal entanglement entropy, namely,
\begin{equation}
 \lim_{\substack{L \to \infty\\f\to 0}}\langle S_A \rangle = L_A \ln 2 \, .
\end{equation}
Here we are in interested in $\langle S_A\rangle$ when $L \to \infty$ for $f > 0$. We define the coefficient of its volume-law term as
\begin{equation} \label{def_sdensity}
 s(f)=\lim_{L\to\infty}\frac{\langle S_A\rangle}{L_A\ln{2}}\,.
\end{equation}\\
In Ref.~\cite{vidmar2018volume} we conjectured that, for quadratic fermionic Hamiltonians with translational invariance, $s(f)$ is a universal function of the subsystem fraction $f$. $s(f)$ is the central object studied in Sec.~\ref{sec:xx-bounds}.

Accordingly, we also define the coefficients of the leading terms of the $m$-th order bounds
\begin{equation}\label{def_spmm}
 s^{\pm}_m(f)=\lim_{L\to\infty}\frac{\langle S^{m\pm}_A\rangle}{L_A\ln{2}}\,,
\end{equation}
which, see Eqs.~(\ref{def_Splus}) and~(\ref{def_Sminus}), depend on
\begin{equation} \label{def_trdensity}
 t_m(f)=\lim_{L\to\infty}\frac{\langle\mathrm{Tr}[\ii \mathbb{J}]_A^{2m}\rangle}{L_A}\,.
\end{equation}
The average $\langle ... \rangle$ in Eq.~(\ref{def_trdensity}) is defined as in Eq.~(\ref{def_Savr}). As we show next, $t_m(f)$ can be computed from averages $\langle N_{\kappa_1}\cdots N_{\kappa_{2m}}\rangle$ over eigenstate occupation numbers.

\subsection{First order bound: Leading term} \label{sec:xy-model-1order}

Here we calculate the leading term in $t_1(f)$ [see Eq.~(\ref{def_trdensity})]. The first step is to calculate the matrix elements of $[\ii \mathbb{J}]$, see Eq.~(\ref{def_J}), for an eigenstate $\ket{\{n_\kappa\}}$. The matrix $[\ii \mathbb{J}]$ can be expressed in terms of $N_\kappa=(-1)^{n_\kappa}$ as
\begin{widetext}
\begin{equation}\label{eq:iJ-for-xx}
 \ii \mathbb{J}\equiv\frac{1}{L}\sum_{{\kappa}\in\mathcal{K}^\pm}N_{\kappa}\left(\begin{array}{c|c} |u_{\kappa}|^2e^{\ii {\kappa}(j -l)}-|v_{\kappa}|^2e^{\ii {\kappa}(l-j)} & u^*_{\kappa}v^*_{\kappa}\left(e^{\ii {\kappa}(l-j)}-e^{\ii {\kappa}(j -l)}\right)\\ \hline u_{\kappa}v_{\kappa}\left(e^{\ii {\kappa}(l-j )}-e^{\ii {\kappa}(j -l)}\right)& -|u_{\kappa}|^2e^{\ii {\kappa}(l-j )}+|v_{\kappa}|^2e^{\ii {\kappa}(j -l)} \end{array}\right)\,.
\end{equation}
\end{widetext}
Here the choice of the set $\mathcal{K}^\pm$, introduced in Eqs.~(\ref{def_Kplus})--(\ref{def_Kminus}), depends on whether an eigenstate $\ket{\{n_\kappa\}}$ contains an even or an odd number of quasiparticles. We first compute the partial average $\langle \mathrm{Tr}[\ii \mathbb{J}]_A^2\rangle_{\pm}$, where we either take only eigenstates with an even ($+$) or an odd ($-$) number of quasiparticles, and then compute the average over all eigenstates
\begin{equation} \label{def_avrtr}
 \langle \mathrm{Tr}[\ii \mathbb{J}]_A^2\rangle=\frac{1}{2}\left(\langle \mathrm{Tr}[\ii \mathbb{J}]_A^2\rangle_++\langle \mathrm{Tr}[\ii \mathbb{J}]_A^2\rangle_-\right).
\end{equation}
It is easy to check that there is an equal number of eigenstates in the even and the odd sectors $\mathcal{H}^\pm$ of the Hilbert space.\footnote{Consider $0=(1-1)^L=\sum^L_{l=0}{L\choose l}(-1)^{l}(1)^{L-l}$. Here, $l$ represents the number of excitations correctly weighted by the binomial coefficient. The positive terms represent the even sectors (even $l$) and the negative ones the odd sector (odd $l$). Clearly, both terms cancel each other, i.e., the even and odd sector both have Hilbert space dimension $2^{L-1}$.} The partial average is defined as
\begin{equation} \label{def_avrpm}
 \langle \mathrm{Tr}[\ii \mathbb{J}]_A^2\rangle_{\pm} = \frac{1}{2^{L-1}} \sum_{\{N_\kappa\}_\pm} {\rm Tr}[\ii \mathbb{J}]_A^2\,,
\end{equation}
where $\{N_\kappa\}_\pm$ corresponds to the sets of all $N_\kappa\in\{1,-1\}$ with $\kappa\in\mathcal{K}^\pm$ and $\prod_{\kappa\in\mathcal{K}^\pm}N_\kappa=\pm 1$. We stress that we compute the average over the eigenstates of $\hat{H}^+_{\mathrm{XY}}$ with an even number of excitations and the average over the eigenstates of $\hat{H}^-_{\mathrm{XY}}$ with an odd number of excitations.

The partial average $\langle \mathrm{Tr}[\ii \mathbb{J}]_A^2\rangle_{\pm}$ can be computed using the technique introduced in Ref.~\cite{vidmar_hackl_17}, which is based on the partial average $\langle N_{\kappa_1}\cdots N_{\kappa_n}\rangle_\pm$. The definition of the latter is analogous to the one in Eq.~(\ref{def_avrpm}). As we are summing over all states, each number $N_{\kappa_i}$ takes values $\pm 1$, subject to the constraint that the total number of $-1$'s is even or odd. However, this constraint does not affect the expectation values $\langle N_{\kappa_1}\cdots N_{\kappa_n}\rangle_\pm$ for $n<L$, which are given by~\cite{vidmar_hackl_17}
\begin{align}
\begin{split}\label{eq:binomial}
 \langle N_{\kappa}\rangle_\pm=0\,,\qquad \langle N_{\kappa_1}N_{\kappa_2}\rangle_\pm=\delta_{\kappa_1\kappa_2}\,,\\ \langle N_{\kappa_1}\cdots N_{\kappa_n}\rangle_\pm=\left\{\begin{array}{ll} 1 & \text{$\kappa_i$ appear in pairs}\\ 0 & \text{else} \end{array}\right.\,.
\end{split}
\end{align}
Only at order $n=L$, these correlation functions are modified due to the above constraint, as explained in Appendix~\ref{app:evenodd}. As we are taking the limit $L\to\infty$, we can use the above expectation values for the bounds $s^{\pm}_m(f)$ at any order $m$ by assuming that $L>m$.

Using Eq.~(\ref{eq:binomial}), one can compute the partial average
\begin{equation} \label{tr2_1order}
 \langle \mathrm{Tr}[\ii \mathbb{J}]_A^2\rangle_{\pm}=2\frac{L_A^2}{L}-\frac{2}{L^2}\sum^{L_A}_{j ,l=1}\sum_{k\in\mathcal{K}^\pm}4|u_\kappa|^2|v_\kappa|^2\cos{2\kappa(j -l)}\,.
\end{equation}
Therefore, taking the total average $\langle \mathrm{Tr}[\ii \mathbb{J}]_A^2\rangle$ defined in Eq.~(\ref{def_avrtr}) is equivalent to summing over all $\kappa\in \mathcal{K}^+\cup\mathcal{K}^-$ and dividing by $2$. Thus, we find
\begin{equation}
 \langle \mathrm{Tr}[\ii \mathbb{J}]_A^2\rangle = 2\frac{L_A^2}{L}-Z(L_A,L) \label{trJ2},
\end{equation}
where
\begin{align}
 Z&(L_A,L) = \nonumber \\ 
 &=\frac{1}{L^2}\sum^{L_A}_{j ,l=1}\sum^{L-1}_{k=1-L}4\,|u_{\pi k/L}|^2|v_{\pi k/L}|^2\cos{\left[\frac{2\pi k(j -l)}{L}\right]}\nonumber\\
 &=\frac{1}{L^2}\sum^{L-1}_{k=1-L}4\,|u_{\pi k/L}|^2|v_{\pi k/L}|^2\,\frac{\sin^2\left(\frac{\pi k L_A}{L}\right)}{\sin^2\left(\frac{\pi k }{L}\right)}\,, \label{Z_long}
\end{align}
in which the sum over $k$ runs only up to $k=L-1$, because $u_{\pi}=0$. We can use
\begin{equation}
 \sum_{k=1-L}^{L-1}\frac{\sin^2\left(\frac{\pi k L_A}{L}\right)}{\sin^2\left(\frac{\pi k }{L}\right)}=2L_A(L-L_A),
\end{equation}
together with the inequality $|u_\kappa|^2|v_\kappa|^2\leq \frac{1}{4}$, because $|u_\kappa|^2+|v_\kappa|^2=1$, to find the bound
\begin{equation}
 0\leq Z(L_A,L)\leq \frac{2L_A(L-L_A)}{L^2}\,.
\end{equation}
This implies that $\lim_{L\to\infty}Z(L_A,L)\leq{\rm const}$. Using Eqs.~(\ref{def_trdensity}) and~(\ref{trJ2}), we arrive at
\begin{equation}\label{eq:t1XY}
 t_1(f) = \lim_{L\to\infty}\left(2\frac{L_A}{L}-\frac{Z(L_A,L)}{L_A}\right)=2f\,.
\end{equation}
This result agrees with the one obtained for translational invariant quadratic models~\cite{vidmar_hackl_17}. Note, however, the XY chain in a transverse field has a nonquadratic boundary term. Our result for this model shows that the boundary term does not change the leading behavior of $t_1$.

\subsection{First order bound: Subleading term} \label{sec:xy-model-1order-sub}

Next, we calculate the first subleading term in $t_1(f)$ in the thermodynamic limit. While it is zero for noninteracting fermions~\cite{vidmar_hackl_17}, it was found for the quantum Ising model ($\gamma=1$) that it is a nonvanishing constant at the critical field $h=J$ and vanishes otherwise~\cite{vidmar2018volume}. Here we generalize that result to arbitrary values of $\gamma$.

The subleading term of the first order bound is obtained by calculating the difference between the exact result $\langle {\rm Tr}[\ii \mathbb{J}]_A^2 \rangle$ and the leading term $2f L_A$. This is exactly $Z(L_A,L)$, defined in Eq.~(\ref{trJ2}). Following Eq.~(\ref{Z_long}), one can write $Z(L_A,L)=\frac{1}{L^2}\sum^{L-1}_{k=1-L}z(\pi k/{L})$ by introducing the function $z(\kappa)$ as
\begin{align}
 z(\kappa)&=4\,|u_{\kappa}|^2|v_{\kappa}|^2\,\sin^2\left(L_A\kappa\right)\,\sin^{-2}\left(\kappa\right)\\
 &=\frac{\gamma^2 J^2\sin^2 (L_A \kappa)}{h^2+2 h J\cos (\kappa)+\left(\gamma ^2-1\right) J^2 \sin^2(\kappa)+J^2}\,.\nonumber
\end{align}
We note immediately that, for $\gamma=0$ (noninteracting fermions), one has $Z(L_A,L)=0$. In order to bound the scaling of $Z(L_A,L)$ as $L\to\infty$ for $\gamma>0$, we need to distinguish two cases:

\begin{figure}[t]
\centering
\noindent\makebox[\linewidth]{\includegraphics[width=\linewidth]{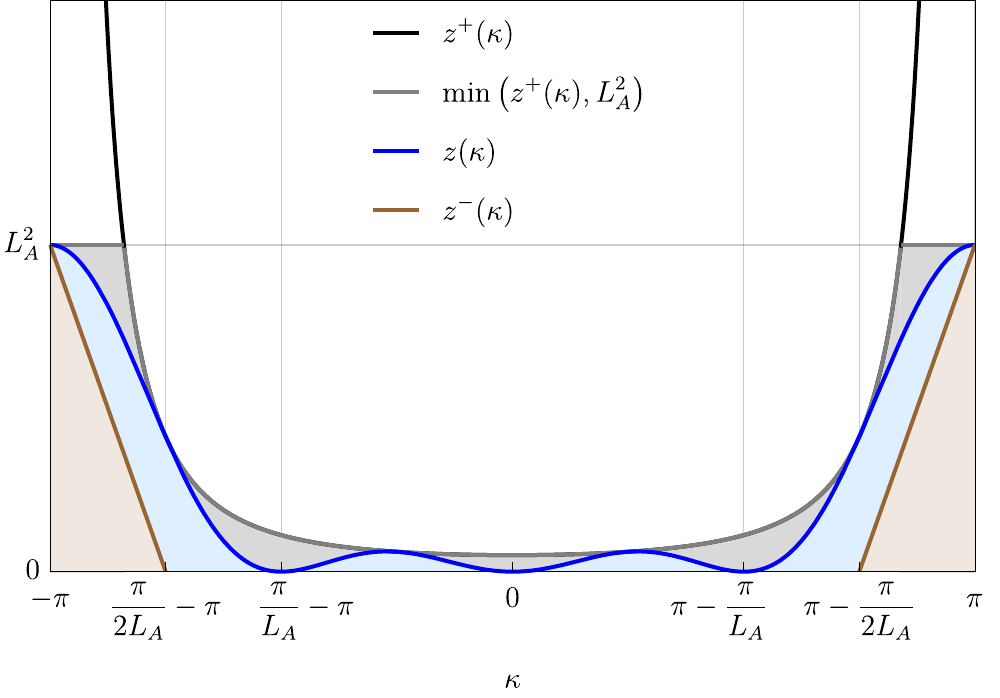}}
\ccaption{Subleading term of the first order bound for the XY model}{Sketch of the function $z(\kappa)$ and its bounds $z^\pm(\kappa)$ for $\gamma=0.9$, $h=J$, $L_A=2$ and $L=4$. We have the inequality $z^-(\kappa)\leq z(\kappa)\leq\min[L_A^2,z^+(\kappa)]\leq z^+(\kappa)$. The integrals over $z^-(\kappa)$ and $\min[L_A^2,z^+(\kappa)]$ corresponding to the shaded surfaces in brown and gray, respectively, both scale as $L_A$, which implies that the same scaling applies to $Z(L_A,L)\sim \frac{1}{\pi L}\int^\pi_{-\pi}d\kappa\,z(\kappa)\sim \mathcal{O}(L_A/L)$, as given by Eq.~(\ref{eq:asymp_z}).}
\label{fig:XY-critical-1st-order-bound}
\end{figure}

\textbf{(i) Critical line ($h=J$):}\\
We first consider the simplest case, $\gamma=1$, for which one can perform the sum analytically
\begin{equation}
 Z(L_A,L)=\frac{1}{L^2}\sum^{L-1}_{k=1-L}\frac{\sin^2 \left(\frac{\pi k L_A}{L}\right)}{2+2 \cos \left(\frac{\pi k}{L}\right)}=\frac{L_A}{L}-\left(\frac{L_A}{L}\right)^2\,,
\end{equation}
where the fact that $L_A$ is an integer allowed us to simplify the summation. This shows that the subleading term $Z(L_A,L)$ does not vanish if $f>0$, and only depends on $f$. For $0<\gamma<1$, we cannot compute the sum analytically, but we can estimate its scaling. Replacing $\sin^2(\kappa L_A)$ by 1 leads to an upper bound
\begin{equation}
 z(\kappa)\leq z^+(\kappa)=\frac{\gamma^2}{2+2\cos (\kappa)+\left(\gamma ^2-1\right) \sin^2(\kappa)}\,,
\end{equation}
which diverges at $\kappa=\pm \pi$. However, the function $z(\kappa)$ does not diverge, it reaches its maximum
\begin{equation}
 z^+(\pm\pi)=L_A^2\,,
\end{equation}
where we used L'Hospital's rule twice. 

To show that, for a fixed $f$, $\lim_{L\to\infty}Z(L_A,L)$ reaches a nonzero value, we approximate $\frac{1}{L}\sum_{k=1-L}^{L-1} z(\pi k/L)\to\frac{1}{\pi}\int^\pi_{-\pi} z(\kappa) d\kappa$. The function $z(\kappa)$ is shown in Fig.~\ref{fig:XY-critical-1st-order-bound}. The scaling of the integral can be predicted by bounding $z(\kappa)$ from below by the triangle function
\begin{equation}
 z^-(\kappa)=\left\{\begin{array}{lcl} L_A^2-2L_A^3\left(1-\frac{|\kappa|}{\pi}\right) & & \pi-\frac{\pi}{2L_A}\leq|\kappa|\leq \pi\\ 0 & & \text{else} \end{array}\right.,
\end{equation}
and from above either by the maximum value $L_A^2$ or by the asymptotic expansion $z^+(\kappa)=(\pi-|\kappa|)^{-2}$ as $|\kappa|\to\pi$. The integral over both functions is proportional to $L_A$. This implies that
\begin{equation}\label{eq:asymp_z}
 Z(L_A,L) \approx \frac{1}{\pi L}\underbrace{\int^\pi_{-\pi}\hspace{-3mm} z(\kappa) \, d\kappa}_{\sim \mathcal{O}(L_A)}\sim\mathcal{O}\left(\frac{L_A}{L}\right)\,,\vspace{0.1cm}
\end{equation}
and proves that $\lim_{L\to\infty} Z(L_A,L) \sim \mathcal{O}(L_A/L)$ for any $\gamma>0$, i.e., that the first subleading correction does not vanish if $f>0$, and only depends on $L_A$ through $f$.

\textbf{(ii) Regime $h\neq J$:}\\
We have $z(\kappa)\leq z^+(\kappa)$ using that $\sin^2(\kappa L_A)\leq 1$, where
\begin{equation}
 z^+(\kappa)=\frac{\gamma^2 J^2}{h^2+2 h J\cos (\kappa)+\left(\gamma ^2-1\right) J^2 \sin^2(\kappa)+J^2}\,.
\end{equation}
$z^+(\kappa)$ takes its maximal values at $\kappa=\pm\pi$, given by
\begin{equation}
 z^+(\pm\pi)=\frac{\gamma^2J^2}{(h-J)^2}\,.
\end{equation}
Therefore
\begin{equation}
 Z(L_A,L)\leq \frac{1}{\pi L}\int^{\pi}_{-\pi}z^+(\kappa)\,d\kappa\leq\frac{1}{L}\frac{2\gamma^2J^2}{(h-J)^2}\,.
\end{equation}
This shows that, irrespective of the value of $f$, the subleading term of the first order bound vanishes in the thermodynamic limit.

\subsection{Average entanglement entropy} \label{sec:xy-model-avr}

Here, we report numerical results for the average eigenstate entanglement entropy $\langle S_A\rangle$ in the spin-1/2 XY chain in a transverse field. They allow us to extract the leading and first subleading terms in the thermodynamic limit, thereby complementing the results for the quantum Ising model ($\gamma=1$) reported in Ref.~\cite{vidmar2018volume}.

\begin{figure*}[!t]
\noindent
\begin{minipage}{.48\textwidth}
\includegraphics[width=\textwidth]{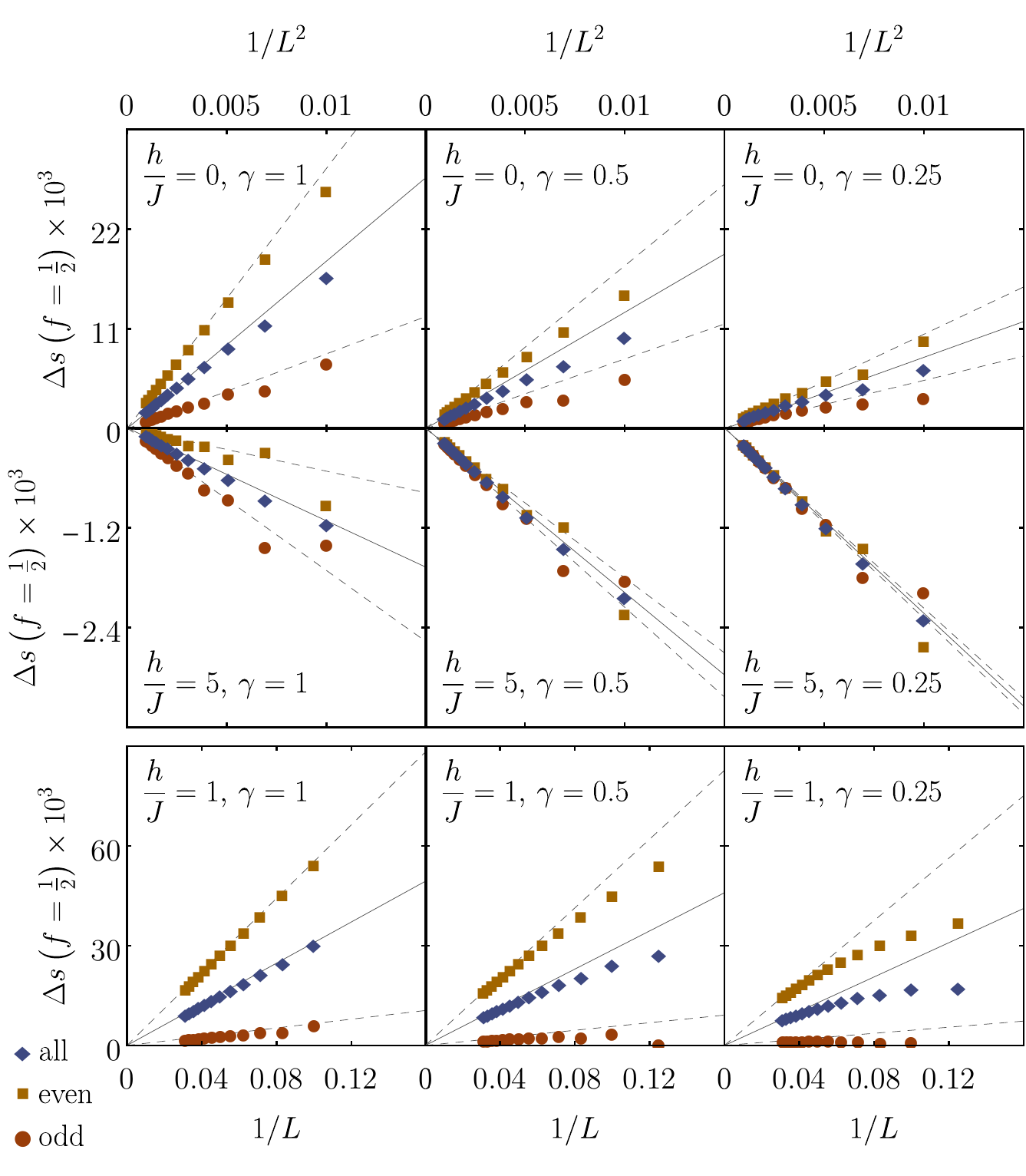}
\end{minipage}
\hspace{.04\textwidth}
\begin{minipage}{.48\textwidth}
\includegraphics[width=\textwidth]{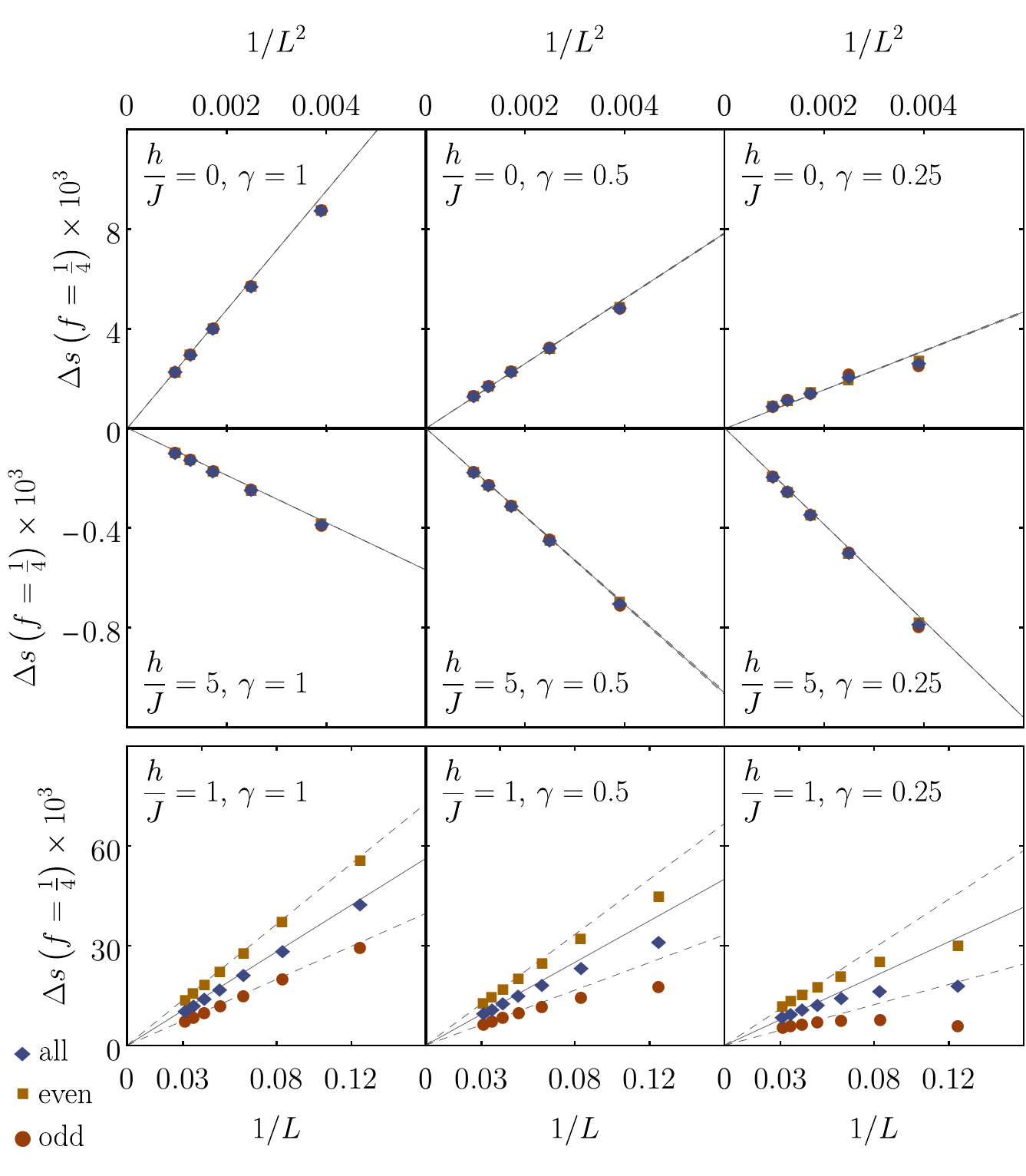}
\end{minipage}
\ccaption{Numerical results for the average entanglement entropy $\langle S_A \rangle$ in the spin-1/2 XY chain in a transverse field}{We show the difference $\Delta s(f)$, defined in Eq.~(\ref{def_dS}), for $f=1/2$ (left panels) and $f=1/4$ (right panels) and different values of $h/J$ and $\gamma$. The results labeled as ``even'' and ``odd'' correspond to $\Delta s(f)_\pm$, respectively, defined in Eq.~(\ref{def_spm}). We find $\Delta s(t) \to 0$ in all cases, suggesting that the leading term of $\langle S_A \rangle$ is universal within the spin-1/2 XY chain in a transverse field. The finite-size scaling analyses reveal that $\Delta s(f)\sim 1/L$ for $h=J$ (i.e., the subleading term in $\langle S_A \rangle$ does not vanish in the thermodynamic limit) and $\Delta s(f)\sim 1/L^2$ otherwise (i.e., the subleading term in $\langle S_A \rangle$ vanishes in the thermodynamic limit).}
\label{fig:Deltas}
\end{figure*}

We define
\begin{equation} \label{def_dS}
 \Delta s(f)=\frac{\langle S_A\rangle}{L_A\ln{2}}-s(f)_{\rm NI} \, ,
\end{equation}
where $\langle S_A\rangle$ is the numerical average in a finite system, and $s(f)_{\rm NI}$ is an extrapolated value of $\langle S_A\rangle/(L_A\ln{2})$ for noninteracting fermions in the thermodynamic limit. For noninteracting fermions, we can accurately obtain the thermodynamic limit value with (at least) a four digit precision because finite-size effects decay exponentially with $L$~\cite{vidmar_hackl_17}. We find $s(f=1/2)_{\rm NI}=0.5378(1)$ and $s(f=1/4)_{\rm NI}=0.7939(8)$.

Figure~\ref{fig:Deltas} shows scaling analyses of $\Delta s(f)$ at $f=1/2$ (left panels) and $1/4$ (right panels), for different values of $\gamma$ and $h/J$. In all cases, we find $\lim_{L\to\infty}\Delta s(f) \to 0$, as for the quantum Ising model~\cite{vidmar2018volume}. This provides further evidence, in addition to the one reported in Ref.~\cite{vidmar2018volume}, of the universality of the leading order term for quadratic fermionic Hamiltonians (and the spin models mappable to them) with translational invariance. 

The finite-size scaling analyses of $\Delta s(f)$ in Fig.~\ref{fig:Deltas} also uncover the nature of the subleading term of $\langle S_A \rangle$. They show that, for $\gamma>0$ and $f>0$, $\Delta s(f)$ scales as $1/L$ at $h=J$ and as $1/L^2$ otherwise. This suggests that, in the spin-1/2 XY chain in a transverse field, $\langle S_A\rangle$ in the thermodynamic limit exhibits a subleading term that does not vanish in the critical line $h=J$ ($\gamma>0$) for $f>0$, and vanishes otherwise. Therefore, as found for the quantum Ising model in Ref.~\cite{vidmar2018volume}, the numerical results for the average and the analytical results for the first order bounds exhibit identical finite-size scalings of the leading and first subleading terms.

In Fig.~\ref{fig:Deltas} we also show results for the entanglement entropy averaged only over eigenstates with an even (or odd) number of quasiparticles. We define [see Eq.~(\ref{def_avrpm})]
\begin{equation} \label{def_spm}
 \Delta s(f)_\pm=\frac{\langle S_A\rangle_\pm}{L_A\ln{2}}-s(f)_{\rm NI} \, .
\end{equation}
The results for $\Delta s(f)_\pm$ in Fig.~\ref{fig:Deltas} suggest that, generically, $\Delta s(f)_+$ and $\Delta s(f)_-$ exhibit a different subleading term. One may wonder what happens if one incorrectly takes the average over all eigenstates using a single set of $\kappa$ values (either $\kappa \in {\cal K}^+$ or $\kappa \in {\cal K}^-$) instead of the correct average involving the true eigenstates, one half of which come from the even sector ($\kappa \in {\cal K}^+$) and the other half of which come from the odd sector ($\kappa \in {\cal K}^-$). Figure~\ref{fig:pcbabc} in Appendix~\ref{app:evenodd} shows that the incorrect calculation produces an error that vanishes $\propto\exp(-\alpha L^\beta)$, where $\alpha$ depends, in general, on the Hamiltonian parameters, the sector used for the average, and on $f$. Numerically, we find that $\beta=1$ for $f=1/4$, and $\beta=1/2$ for $f=1/2$.

\section{Universal volume-law coefficient}\label{sec:xx-model}

We focus next on tightening the bounds for the volume-law coefficient $s(f)$ defined in Eq.~(\ref{def_sdensity}), which is a concave function of the subsystem fraction $f$ that was conjectured to be universal for quadratic fermionic Hamiltonians with translational invariance~\cite{vidmar2018volume}. In Sec.~\ref{sec:u-conjecture}, we provide details about the universality conjecture for $s(f)$. In Sec.~\ref{sec:xx-bounds}, we report the fourth order bounds for $s(f)$ for noninteracting fermions.

\subsection{Conjecture about universality}\label{sec:u-conjecture}

In Ref.~\cite{vidmar2018volume}, we reported numerical results supporting the conjecture that the volume-law coefficient $s(f)$ is universal for quadratic fermionic Hamiltonians with translational invariance. That conjecture is further supported by our earlier results here for the spin-1/2 XY chain in a transverse field, for which $s(f)$ was found to be identical to that of noninteracting fermions independently of the values of $\gamma$, $h$, and $f$ chosen. 

We should stress that this universality cannot include \emph{all} quadratic fermionic Hamiltonians. A simple counterexample is given by the ultra-local Hamiltonian
\begin{equation}
\hat{H} = \sum^L_{j=1}\epsilon_j\hat{f}^\dagger_j \hat{f}_j^{} +E_0\,,
\end{equation}
with randomly selected $\epsilon_j$'s. Every eigenstate of this model has zero entanglement entropy. Another counterexample involving randomness is the quadratic part of the Sachdev-Ye-Kitaev model, for which the average eigenstate entanglement entropy was computed analytically~\cite{liu_chen_18} and differs from our numerical results in Ref.~\cite{vidmar_hackl_17} for translationally invariant quadratic systems.

The most general translationally invariant quadratic Hamiltonian in a one-dimensional lattice with periodic boundary conditions, which can be generalized to higher dimensions in a straightforward way, can be written as
\begin{equation}\label{eq:Hquadratic}
\hat{H}=-\sum^L_{j,l =1} \left(\Delta_l \hat{f}_j ^\dagger \hat{f}_{j+l}^\dagger+\Delta_l^* \hat{f}^{}_j \hat{f}^{}_{j+l}+t_l \hat{f}^\dagger_j \hat{f}_{j+l}^{}+t_l^* \hat{f}^\dagger_{j+l}\hat{f}_{j}^{}\right),
\end{equation}
with $j,l\in\mathbb{Z}_L$, namely, $j+L\equiv j$. Hamiltonian~\eqref{eq:Hquadratic} can be diagonalized by first applying a Fourier transformation and then performing a two mode squeezing. Explicitly, we define
\begin{equation}
\hat{c}_\kappa^{} = \frac{1}{\sqrt{L}} \sum_{\kappa\in\mathcal{K}} e^{\ii\kappa j} \hat{f}_j^{} \quad \text{and} \quad \hat{c}^\dagger_\kappa = \frac{1}{\sqrt{L}} \sum_{\kappa\in\mathcal{K}} e^{-\ii\kappa j} \hat{f}^\dagger_j \,,
\end{equation}
with $\mathcal{K}=\left\{\frac{2\pi k}{L}\,\big|\,k\in\mathbb{Z}\,,-\frac{L}{2} < k \leq \frac{L}{2}\right\}$, giving rise to the Fourier transformed Hamiltonian
\begin{equation}
\hat{H} = -\sum_{\kappa\in\mathcal{K}} \left[ \tilde{\Delta}_\kappa \hat{c}_\kappa^\dagger \hat{c}_{-\kappa}^\dagger + \tilde{\Delta}^*_\kappa \hat{c}_\kappa ^{} \hat{c}_{-\kappa}^{} + (\tilde{t}_\kappa + \tilde{t}^*_{\kappa}) \hat{c}_\kappa^\dagger \hat{c}_{\kappa}^{}\right],
\end{equation}
with the parameters
\begin{equation}
\tilde{\Delta}_\kappa=\sum_{j}e^{\ii\kappa j}\Delta_j,\quad\text{and}\quad\tilde{t}_\kappa=\sum_{j}e^{-\ii\kappa j}t_j \,.
\end{equation}

For the two mode squeezing, one determines the parameters $u_\kappa$ and $v_\kappa$ so that $\hat{\eta}_\kappa^{} = u_\kappa\hat{c}_\kappa^{} - v_\kappa^*\hat{c}_{-\kappa}^\dagger$ leads to the quadratic Hamiltonian
\begin{equation}
\hat{H}=\sum_{\kappa\in\mathcal{K}} \epsilon_\kappa\, \hat{\eta}_\kappa^\dagger \hat{\eta}_\kappa^{} + E_0\,,
\end{equation}
where $E_0$ is an overall constant. Note that $u_\kappa$ and $v_\kappa$ satisfy the conditions
\begin{equation}
u_\kappa=u_{-\kappa}\, ,\quad\text{and}\quad v_\kappa=-v_{-\kappa}\,,
\end{equation}
which follow from the fact that one squeezes the mode pairs $(\kappa,-\kappa)$. One always has $u_0=1$ and $v_0=0$, and, for even $L$, one also has $u_{\pi}=1$ and $v_{\pi}=0$. 

In order to test the universality of the volume-law coefficient $s(f)$ for translationally invariant quadratic Hamiltonians, in Ref.~\cite{vidmar2018volume} we studied a class of Hamiltonians defined by choosing the parameters $u_\kappa$ and $v_\kappa$ to be simple continuous and periodic functions on the interval $[0,\pi]$.\footnote{A natural choice of \emph{regular} limit functions are those continuous on the circle, i.e., periodic such that $u_{-\pi}=u_{\pi}$ and $v_{-\pi}=v_{\pi}=0$. It is conceivable that one may also need to require additional regularity conditions on $u_\kappa$ and $v_\kappa$. This is challenging to verify numerically and remains to be done.} Finite-size scaling analyses showed that, in the thermodynamic limit, $s(f)$ in those Hamiltonians becomes identical to that of noninteracting fermions.

Based on those results, we conjectured that the volume-law coefficient $s(f)$ is universal for quadratic fermionic Hamiltonians with translational invariance~\cite{vidmar2018volume}. The conjecture can be formulated as follows.

{\bf Universality conjecture:}
For translationally invariant quadratic fermionic Hamiltonians in arbitrary dimensions, with $V$ sites and a regular subsystem region consisting of $V_A$ sites, the volume-law coefficient of the average (over all eigenstates) eigenstate entanglement entropy is a universal function in the thermodynamic limit:
\begin{equation}
 s(f)=\lim_{V\to\infty}\frac{\langle S_A\rangle}{V_A\ln{2}},\quad\text{with}\quad f=\frac{V_A}{V}\,.
\end{equation}
In the case of degenerate eigenspaces, the basis for the average should be that obtained in the limit of vanishingly small translationally invariant perturbations that break all degeneracies.

We also conjecture that $s(f)$ is typical for the entanglement entropy of eigenstates, which implies that:
\begin{equation} \label{def_variance}
 \Sigma_s(f)=\lim_{L\to\infty}\frac{\langle S_A^2\rangle-\langle S_A\rangle^2}{(V_A\ln{2})^2}=0.
\end{equation}

A few comments are in order about this conjecture.

(i) \textit{Translational invariance} refers to a discrete symmetry of a regular lattice in $D$ dimensions. For definiteness, one can consider a hypercubic lattice with periodic boundary conditions and $L$ sites along each principal direction, so that the total number of sites is $V=L^D$. This is the setup used to prove the universal first order bound with a Hamiltonian that is invariant under discrete lattice translations~\cite{vidmar_hackl_17}.

(ii) \textit{Regular subsystem region} refers to subsets consisting of $V_A$ sites that have a well-defined geometry in the thermodynamic limit. This intuition stems from our proof for the universality of the first order bound~\cite{vidmar_hackl_17}. Even though we only considered subsystems defined as cuboids of length $L_d$ along the $d$-th dimension, we expect that the results do not change if one deforms these regions as long as they have a well defined thermodynamic limit. This excludes fractal subsystems or other definitions that select subsystem sites based on some nongeometric criterion. The reason for this condition lies in the fact that already at the level of the first order bound, the generalization of Eq.~(\ref{tr2_1order}) to arbitrary dimensions yields
\begin{equation}
	\langle \mathrm{Tr}[\ii \mathbb{J}]^2_{A}\rangle=\frac{2V_{A}^2}{V}-\sum_{l=1}^V\sum_{i,j\in {A}}\frac{8|v_{\vec{\kappa}_l}|^2|u_{\vec{\kappa}_l}|^2\cos{2\vec{\kappa}_l(\vec{x }_i-\vec{x }_j)}}{V^2}\,,
\end{equation}
where one could select the points $\vec{x }_i$ in the subsystem as a function of the total system size to get specific contributions from the cosine function that would violate our bound. This is not possible if the subsystem approaches a regular geometric region in the thermodynamic limit.

(iii) \textit{Typicality} is an additional part of the conjecture stating that the average volume-law coefficient $s(f)$ is typical for energy eigenstates. This means that, in the thermodynamic limit, all but a  measure zero of the energy eigenstates have a leading term of the entanglement entropy described by $L_A \ln 2 \, s(f)$. In Ref.~\cite{vidmar_hackl_17} we proved this for noninteracting fermions by showing that the variance $\Sigma_s(f)$, defined in Eq.~(\ref{def_variance}), vanishes at least as $1/\sqrt{L}$ with increasing system size.

\subsection{Noninteracting fermions} \label{sec:xx-bounds}

Motivated by the conjecture discussed in Sec.~\ref{sec:u-conjecture}, and in order to learn more about $s(f)$ for quadratic fermionic Hamiltonians with translational invariance, we tighten the bounds reported in Ref.~\cite{vidmar_hackl_17} for noninteracting fermions in one dimension [Eq.~(\ref{def_Hfree})]. The single-particle eigenstates of that model are plane waves, so this is the model for which the analytic calculation of the bounds is easiest.

\subsubsection{Expansion up to fourth order} \label{sec:expansion}

In the basis given by $\hat\xi^a\equiv(\hat f_1,\cdots,\hat f_L, \hat f_1^\dagger,\cdots, \hat f_L^\dagger)$, the matrix $\ii \mathbb{J}$, see Eq.~(\ref{def_J}), is block diagonal and explicitly given by
\begin{equation}
\ii \mathbb{J} \equiv \sum_{\kappa\in\mathcal{K}} N_\kappa \left(\begin{array}{c|c}
\ee^{\ii\kappa(j - l)} & 0\\
\hline
0 & -\ee^{\ii\kappa(j - l)}
\end{array}\right),
\end{equation}
with $\mathcal{K}=\left\{\frac{2\pi k}{L}\,\big|\,k\in\mathbb{Z}\,,-\frac{L}{2} < k \leq \frac{L}{2}\right\}$.

As we proved in Ref.~\cite{vidmar_hackl_17}, the coefficients of the leading term of the first order bounds are universal for translationally invariant quadratic systems
\begin{equation}
s^+_1(f)=1-\frac{f}{2\ln{2}}\,\,,\quad\text{and}\quad
s^-_1(f)=1-\frac{f}{\ln{2}}\,.
\end{equation}
The same results follow for the spin-1/2 XY model in the presence of a transverse field, after substituting $t_1(f)$ from Eq.~\eqref{eq:t1XY} in Eqs.~\eqref{def_Splus} and~\eqref{def_Sminus}. 

Computing higher order bounds becomes increasingly complicated, and there are several subtleties to note. We develop a general procedure to compute averages of traces of $[\ii \mathbb{J}]_{A}^{2n}$ based on generalized Wick contractions. This procedure is discussed in Appendix~\ref{app1}. The main insight from our analysis is that the term $\langle {\rm Tr} [\ii \mathbb{J}]_{A}^{2n} \rangle/L_A$ is a polynomial that, when $L\to \infty$ and $f\to 0$, only contains powers from $f^{n}$ to $f^{2n-1}$. Table~\ref{tab:expansionTrJA} lists the coefficients of $f^{n}$ to $f^{2n-1}$ in $t_n(f)$, defined in Eq.~(\ref{def_trdensity}), for $n\leq4$ (see Appendix~\ref{app1} for the corresponding calculations).

\begin{table}[!h]
\ccaption{Coefficients in $t_n$}{List of the coefficients $\alpha_n^{(m)}$ in $t_n(f)=\sum^{2n-1}_{m=n}\alpha_n^{(m)}\,f^{m}$, defined in Eq.~(\ref{def_trdensity}), for $n\leq4$.}
\label{tab:expansionTrJA}
\begin{center}
\begin{tabular}{c||c|c|c|c|c|c|c}
$ t_n$ & \hspace{2mm}$\alpha_n^{(1)}$\hspace{2mm} & \hspace{2mm}$\alpha_n^{(2)}$\hspace{2mm} & \hspace{2mm}$\alpha_n^{(3)}$\hspace{2mm} & \hspace{2mm}$\alpha_n^{(4)}$\hspace{2mm} & \hspace{2mm}$\alpha_n^{(5)}$\hspace{2mm} & \hspace{2mm}$\alpha_n^{(6)}$\hspace{2mm} & \hspace{2mm}$\alpha_n^{(7)}$\hspace{2mm} \\[1mm] \hline &&&&&&&\\[-3mm]
$n=1$ & $2$ & & & & & & \\
$n=2$ & & $\frac{16}{3}$ & $-4$ & & & & \\
$n=3$ & & & $22$ & $-48$ & $32$ & & \\
$n=4$ & & &  & $\frac{1816}{15}$ & $-\frac{2592}{5}$ & $\frac{54640}{63}$ & $-544$
\end{tabular}
\end{center}
\end{table}

\begin{figure}[!h]
\includegraphics[width=1.05\linewidth]{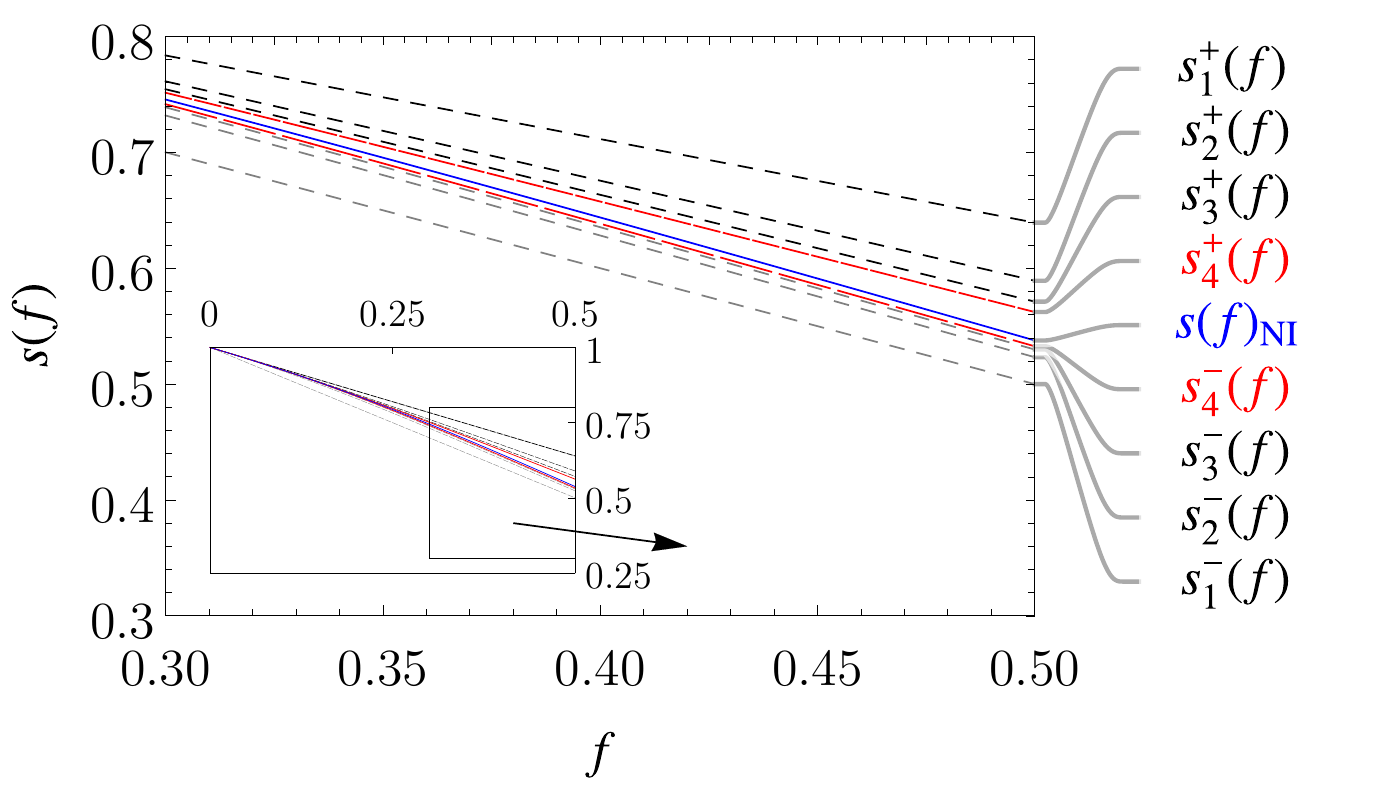}
\vspace{-5mm}
\ccaption{Upper and lower bounds of the volume-law coefficient from first to fourth order}{Bounds $s^\pm_n(f)$, and the numerically calculated average $s(f)_{\rm NI}$ for noninteracting fermions, as functions of the subsystem fraction $f\in[0.3,0.5]$. The inset shows the full range $f\in[0,0.5]$. The results for noninteracting fermions were obtained for a chain with $L=36$ sites and periodic boundary conditions~\cite{vidmar_hackl_17}. }
\label{fig:one-to-fourth-order-bounds}
\end{figure}

Using $t_n(f)$, one can write the volume-law coefficients of higher order bounds $s^\pm_n(f)$ [see Eqs.~\eqref{def_Splus},~\eqref{def_Sminus}, and~\eqref{def_spmm}] as
\begin{align}\label{eq:snp}
s^+_n(f)&=1-\sum^n_{m=1}\frac{t_m(f)}{4m(2m-1)\ln{2}}\,,\\
s^-_n(f)&=s^+_{n-1}(f)-t_n(f)\left(\sum^\infty_{m=n}\frac{1}{4m(2m-1)\ln{2}}\right)\,.\label{eq:snm}
\end{align}
Substituting the coefficients from Table~\ref{tab:expansionTrJA} in Eqs.~\eqref{eq:snp} and~\eqref{eq:snm}, we find the following fourth order bounds
\begin{widetext}
\begin{align} \label{def_sp4}
 s^+_4(f)&=1-\frac{1}{\ln{2}}\left[\frac{f}{2}+\frac{2 f^2}{9}+\frac{f^3}{5}+\frac{59 f^4}{210}-\frac{86 f^5}{21}+\frac{3415 f^6}{441}-\frac{34 f^7}{7}+\frac{1}{112}t_4^{\mathrm{cor}}(f)\right]\,,\\
 \begin{split}
 s^-_4(f)&=1-\frac{1}{\ln{2}}\Bigg[\frac{f}{2}+\frac{2 f^2}{9}+\frac{f^3}{5}+f^4 \left(\frac{908 \ln{2}}{15}-\frac{8579}{225}\right)+f^5 \left(\frac{12028}{75}-\frac{1296 \ln{2}}{5}\right)\\
 &\left.\qquad\qquad\quad+f^6 \left(\frac{27320 \ln{2}}{63}-\frac{50542}{189}\right)+f^7 \left(\frac{2516}{15}-272 \ln{2}\right)+t_4^{\mathrm{cor}}(f)\sum^\infty_{m=4}\frac{1}{4m(2m-1)}\right]\,, \label{def_s4m}
 \end{split}
\end{align}
\end{widetext}
where the function $t_4^{\mathrm{cor}}(f)$ is a piecewise defined polynomial, see Eq.~(\ref{def_tcorr}) and its derivation in Appendix~\ref{ssec:fourth-order}. $t_4^{\mathrm{cor}}(f)$ arises because of a subtle nonanalytic contribution in the computation of the traces, which appears at fourth order for $f\geq \frac{1}{4}$. In Fig.~\ref{fig:one-to-fourth-order-bounds}, we plot all the volume-law coefficients of the bounds $s_n^\pm(f)$, from the first order to the fourth order, along with the numerically calculated volume-law coefficient of the average $s(f)_{\rm NI}$. The bounds can be seen to be very tight, specially $s_4^-(f)$.

\subsubsection{Asymptotic analysis}

As mentioned in Sec.~\ref{sec:expansion}, and as discussed in Appendix~\ref{ssec:fourth-order} and Ref.~\cite{vidmar_hackl_17},
\begin{equation}
 t_n(f)=\sum^{2n-1}_{m=n}\alpha_n^{(m)}\,f^{m}\quad\text{for}\quad 0\leq f\leq \frac{1}{n}\,,
\end{equation}
where the requirement $f\in[0,\frac{1}{n}]$ ensures that nonanalytic contributions of the type $t_n^{\mathrm{cor}}(f)$, seen already in Eqs~\eqref{def_sp4} and~\eqref{def_s4m}, are absent. Using this, if one looks into Eqs.~\eqref{eq:snp} and~\eqref{eq:snm}, one can see that the volume-law coefficient of the upper bound $s_n^+(f)$ describes the exact asymptotic expansion of $s(f)$ through order $f^n$. 

Therefore, the volume-law coefficient of the upper bound $s_4^+(f)$, see Eq.~(\ref{def_sp4}), describes the exact asymptotics of $s(f)$ through order $f^4$
\begin{equation} \label{def_sasy}
 s_{\rm asy}(f) = 1 - \frac{1}{\ln 2} \left[ \frac{f}{2} + \frac{2f^2}{9} + \frac{f^3}{5} + \frac{59 f^4}{210} + {\cal O}(f^5) \right] \, .
\end{equation}

We plot $s_{\rm asy}(f)$ in Fig.~\ref{fig:intro}, and compare it to the numerical results for $s(f)_{\rm NI}$ and to the fourth order bounds, $s_4^\pm(f)$, reported in Eqs.~(\ref{def_sp4}) and~(\ref{def_s4m}). The agreement between all the curves is remarkable up to $f\approx0.3$.

The fact that, at order $n$, we generically find nonanalytic behavior for $f>\frac{1}{n}$ leads to the question of whether the limit function is nonanalytic. We are actually not aware of any simple analytic function whose asymptotic expansion around $f\to 0$ agrees with the terms computed so far. On the other hand, there is the possibility that the nonanalytic parts cancel each other in the total sum, so that question remains open at this point.

\begin{figure*}[!t]
\noindent
\begin{minipage}{.48\textwidth}
\includegraphics[width=\textwidth]{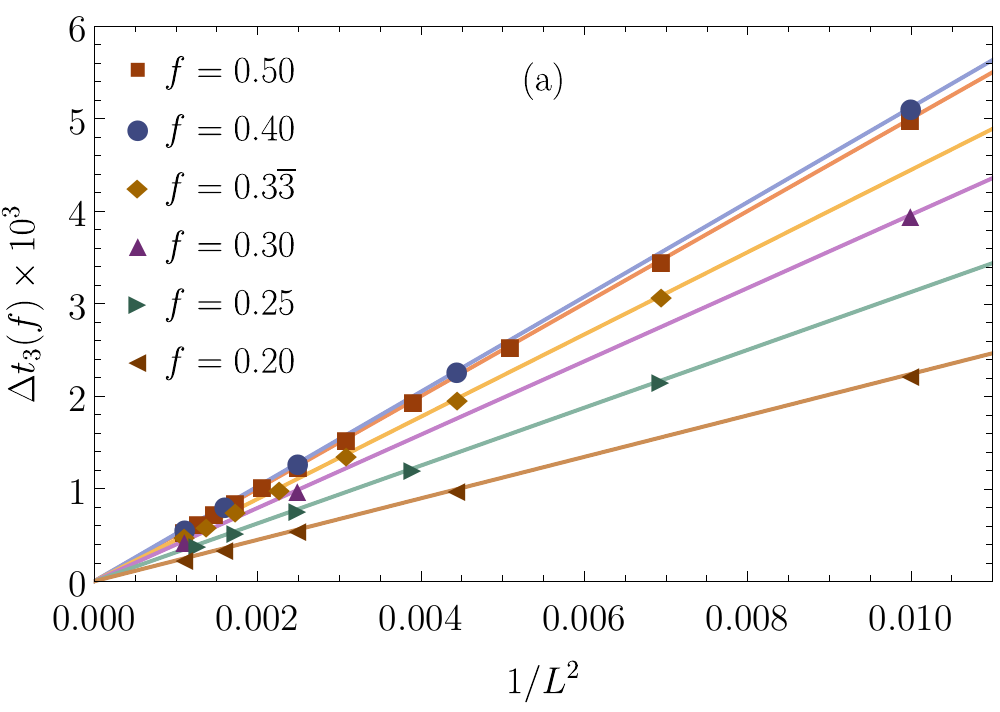}
\end{minipage}
\hspace{.04\textwidth}
\begin{minipage}{.48\textwidth}
\includegraphics[width=\textwidth]{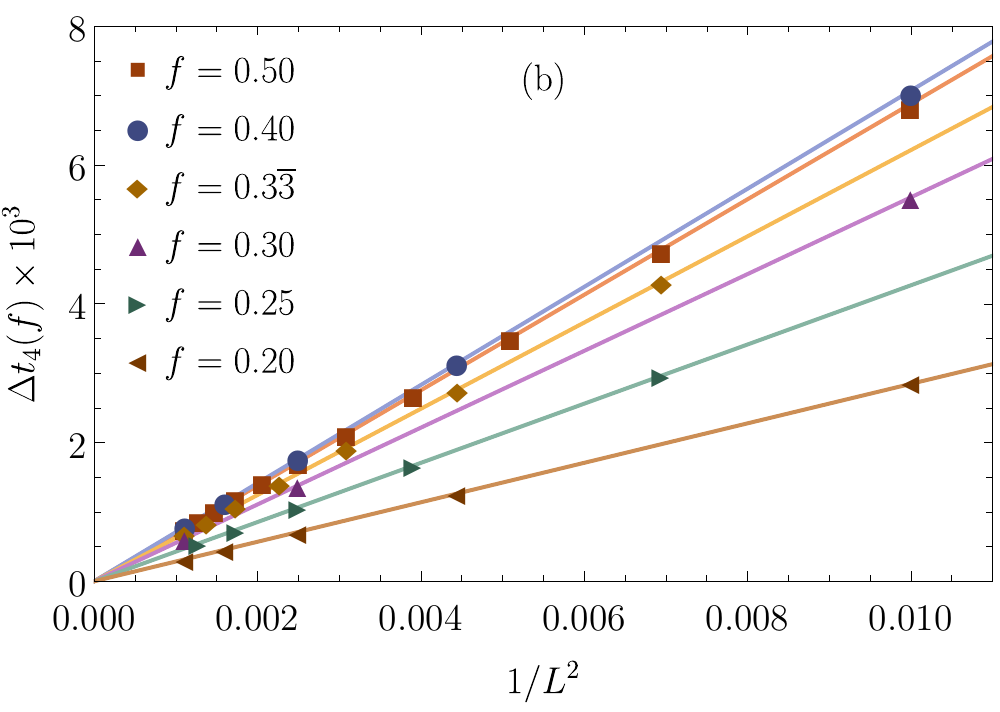}
\end{minipage}
\ccaption{Scaling analysis of the difference between the numerical and the analytical results}{Difference $\Delta t_n(f)$ [see Eq.~\eqref{eq:diftn}] for (a) $n=3$ and (b) $n=4$ and different values of $f$. $\mathrm{Tr} \langle[\ii \mathbb{J}]^{2n}\rangle/L_A$ is calculated numerically for noninteracting fermions in finite chains with $L$ sites and periodic boundary conditions.}
\label{fig:Deltatn}
\end{figure*}

In light of the cumbersome expressions for the volume-law coefficients of the upper and lower bounds up to fourth order, we test their correctness by comparing our analytical results to those obtained using exact numerical calculations. We compute $\langle \mathrm{Tr}[\ii \mathbb{J}]_A^{2n}\rangle/L_A$ numerically in finite systems, and then study the scaling of the difference between the numerical and the analytical (from Table~\ref{tab:expansionTrJA}) results with increasing the system size. Figure~\ref{fig:Deltatn} shows that the difference
\begin{equation}\label{eq:diftn}
 \Delta t_n(f) = \frac{\mathrm{Tr} \langle[\ii \mathbb{J}]^{2n}\rangle}{L_A}-t_n(f)
\end{equation}
vanishes as $1/L^2$ both for $t_3(f)$ and $t_4(f)$. Of particular importance is the agreement at fourth order, which depends on the piecewise defined nonanalytic part $t_4^{\mathrm{cor}}(f)$ for $f>\frac{1}{4}$.

\section{Summary and Discussion} \label{sec:conclusion}

In Sec.~\ref{sec:xy-model}, we discussed analytical results for the first order bounds of the average eigenstate entanglement entropy $\langle S_A\rangle$ for the translationally invariant spin-1/2 XY chain in a transverse field. Those bounds were found to exhibit leading terms that scale linearly with the subsystem volume, which are independent of the model parameters ($J$, $\gamma$, and $h$), and that are identical to the ones obtained for noninteracting fermions. The leading terms of the volume-law coefficients of the bounds only depend on the subsystem fraction $f$. On the other hand, in the thermodynamic limit, the subleading terms of the bounds were found to be nonzero in the critical line $h=J$ ($\gamma>0$) for $f>0$, and vanish otherwise. The same behavior was obtained numerically for the average eigenstate entanglement entropy. These results generalize our findings in Ref.~\cite{vidmar2018volume} for the quantum Ising chain, which corresponds to $\gamma=1$ in Eq.~\eqref{eq:HXY1}.

In Sec.~\ref{sec:xx-model}, motivated by the conjectured universality of the leading order term of the average eigenstate entanglement entropy of translationally invariant quadratic fermionic Hamiltonians (see Sec.~\ref{sec:u-conjecture}), we tightened the bounds for the volume-law coefficient $s(f)$ for noninteracting fermions. $s(f)$ is a concave function of the subsystem fraction $f$, and we computed up to its fourth order bounds. For this, we developed a series expansion method that can be applied to arbitrarily high orders in the expansion of $s(f)$. About $f=0$, our results determine the series expansion of the volume-law coefficient $s(f)$ through $f^4$. Our analytical calculations reduce to the determination of combinatoric factors, which can be found algorithmically using the diagrammatic representation of the generalized Wick's theorem presented in Appendix~\ref{app:Wick-diagrams}. 

Intriguingly, we found that nonanalytic contributions arise in volume-law coefficients that provide bounds for $s(f)$. The nonanalytic contributions are piecewise defined polynomials of $f$, and the series expansion allows us to compute them in each piecewise range of the subsystem fraction $f$. The piecewise polynomial corrections to $s(f)$ for $f>\frac{1}{n}$ (where $n$ is the order of the series expansion) suggest that the function $s(f)$ may not be analytic. In principle, about $f=0$, one can find the expansion in $f$ to arbitrarily high order, but for every interval $[0,\epsilon]$, there exists an $n$ with $n>\frac{1}{\epsilon}$, such that a nonanalytic correction appears. We cannot exclude the possibility that these piecewise defined terms ultimately cancel each other in the total sum, so the question of whether $s(f)$ is an analytic function of $f$ remains open. We should emphasize though that, even at fourth order, the piecewise correction represents an almost negligible change to the function value.

\begin{acknowledgements}
We thank J. Eisert and I. Roth for discussions. We acknowledge support from a Mebus Fellowship (L.H.), the Max Planck Harvard Research Center for Quantum Optics (L.H.), the Slovenian Research Agency research core funding No.~P1-0044 (L.V.), and the National Science Foundation Grant Nos.~PHY-1707482 (M.R.) and~PHY-1806428 (E.B.). The computations were done at the Institute for CyberScience at Penn State.
\end{acknowledgements}

\appendix

\section{Diagonalization of the XY model}\label{app:XYmodel}
For the XY model introduced in Eq.~(\ref{eq:HXY1}), we explain in detail the required transformations that lead to Eq.~(\ref{eq:HXYn}). We follow closely Ref.~\cite{vidmar16} and draw attention to subtleties arising from the boundary terms.

First, expressing the spin operators $\hat{S}_j ^{\mathrm{X}}$ and $\hat{S}_j ^{\mathrm{Y}}$ in terms of ladder operators $\hat{S}^\pm_j = \hat{S}_j ^{\mathrm{X}}\pm\ii\hat{S}_j ^{\mathrm{Y}}$. Then, performing a Jordan-Wigner transformation by writing all spin operators in terms of fermionic creation and annihilation operators $\hat{f}_j^\dagger$ and $\hat{f}_j^{}$, namely,
\begin{equation}
\hat{S}_j ^{+}=\hat{f}_j ^\dagger \,\exp\left(-\ii \pi\sum^{j-1}_{l=1}\hat{f}^\dagger_l\hat{f}_l^{}\right)\,,
\end{equation}
and $\hat{S}^{\mathrm{Z}}_j=\hat n_j - 1/2$, where $\hat n^{}_j = \hat f_j^\dagger\hat f^{}_j$. Not only does this transformation provide an isomorphism between operators, but also preserves the bipartite entanglement, namely, the entanglement entropy associated to a region consisting of adjacent sites $(1,\,2,\ldots,j)$ is the same regardless of whether one uses the tensor product structure generated by the spin operators or by the fermionic creation and annihilation operators. This is a consequence of the remarkable fact that, despite the Jordan-Wigner transformation being nonlocal, the operator $\hat{S}^\pm_j$ only depends on the creation and annihilation operators $\hat{f}^\dagger_l$ and $\hat{f}_l^{}$ in the range $1\leq l\leq j$. The Hamiltonian expressed in terms of the fermionic operators is, up to a constant,
\begin{align}
\begin{split}
\hat{H}_{\mathrm{XY}}=&-\frac{J}{2}\sum^L_{j=1}\left[\hat{f}_j^\dagger(\hat{f}_{j+1}^{}+\gamma \hat{f}_{j+1}^\dagger)+\text{H.c.}\right] -h \hat{N} \\
&+\frac{J}{2}\left[\hat{f}_1^\dagger(\hat{f}_{L}^{}+\gamma \hat{f}_{L}^\dagger)+\text{H.c.}\right](\hat{P}+1)\,,
\label{eq:HP}
\end{split}
\end{align}
where $\hat{P}=e^{\ii \pi\hat{N}}$ is the parity operator, and $\hat{N}=\sum^L_{j=1}\hat n_j$. The last term in Eq.~\eqref{eq:HP} is a boundary term. 

Because of the operator $\hat{P}$, the Hamiltonian $\hat H_{\rm XY}$ in Eq.~(\ref{eq:HP}) is not quadratic in $\hat{f}_j^\dagger$ and $\hat{f}_j$. The nonquadratic term containing $\hat{P}$ distinguishes the sectors with even and odd eigenvalues of the number operator $\hat{N}$. To address this, the Hilbert space can be decomposed as direct sum $\mathcal{H}=\mathcal{H}^+\oplus\mathcal{H}^-$, where $\mathcal{H}^+$ and $\mathcal{H}^-$ are the eigenspaces of the number parity operator $\hat{P}=e^{\ii\pi\hat{N}}$ with eigenvalues $\pm 1$. The projectors onto these eigenspaces are given by
\begin{equation} \label{def_proj}
\hat{\mathcal{P}}^\pm=\frac{1}{2}(\mathds{1}\pm \hat{P})\,.
\end{equation} 

From here, we obtain exact eigenstates of $\hat{H}_{\mathrm{XY}}$ in both sectors. We diagonalize $\hat{H}_{\mathrm{XY}}$ over $\mathcal{H}^\pm$ individually by applying the Fourier transformations
\begin{equation}
\hat{c}_\kappa=\frac{1}{\sqrt{L}}\sum^L_{j =1}e^{\ii \kappa j }\hat{f}_j \,,
\end{equation}
where $\kappa\in\mathcal{K}^\pm$ introduced in Eq.~(\ref{def_Kminus}). The resulting Hamiltonian $\hat{H}_{\mathrm{XY}} = \hat{H}_{\mathrm{XY}}^+ \hat{\mathcal{P}}^+ + \hat{H}_{\mathrm{XY}}^- \hat{\mathcal{P}}^-$ is composed of the quadratic pieces
\begin{equation} \label{def_Hpm}	
\hat{H}_{\mathrm{XY}}^\pm=\sum_{\substack{\kappa\in\mathcal{K}^\pm }} \left[a_\kappa\left(\hat{c}_\kappa^\dagger\hat{c}_\kappa - \frac{1}{2}\right)-\ii b_\kappa\left(\frac{\hat{c}_\kappa^\dagger\hat{c}_{-\kappa}^\dagger-\hat{c}_{-\kappa}\hat{c}_{\kappa}}{2}\right)\right],
\end{equation}	
with
\begin{equation}
a_\kappa=-J\cos(\kappa)-h\,,\quad\text{and}\quad b_\kappa=\gamma\, J\sin(\kappa)\,.
\end{equation}
At this point, one only needs to perform individual fermionic two-mode squeezing transformations mixing the mode pairs $(\kappa,-\kappa)$, with the form
\begin{equation}\label{eq:eta-from-c}
\hat{\eta}_\kappa= u_\kappa \hat{c}_\kappa - v_\kappa^* \hat{c}_{-\kappa}^\dagger\,.
\end{equation}
The transformation coefficients are
\begin{align}
\begin{split}
u_\kappa&=\frac{\varepsilon_\kappa+a_\kappa}{\sqrt{2\varepsilon_\kappa(\varepsilon_\kappa+a_\kappa)}}\,,\quad v_\kappa=\frac{\ii b_\kappa}{\sqrt{2\varepsilon_\kappa(\varepsilon_\kappa+a_\kappa)}} \,,\\
\varepsilon_\kappa&=\sqrt{h^2+2h J\cos(\kappa)+J^2+(\gamma^2-1)J^2\sin(\kappa)^2}\,.	\hspace{-2em}{}
\end{split}
\end{align}
As fermionic Bogoliubov coefficients, $u_\kappa$ and $v_\kappa$ satisfy $|u_\kappa|^2+|v_\kappa|^2=1$. The cases $\kappa=0$ and $\pi$, for which the Hamiltonian is already diagonal, are already included if one chooses the limits of $u_\kappa$ and $v_\kappa$ consistently. One has $u_0=0$ and $v_0=\ii$, while for $\kappa=\pi$, one needs to distinguish
\begin{equation}
u_\pi=\begin{cases} 0 & h>J\\ 1 & h<J \end{cases}, \quad\text{and}\quad v_\pi=\begin{cases}\ii & h>J\\ 0 & h<J \end{cases}\,.
\end{equation}
(For $h=J$, $\hat{c}_\pi$ and $\hat{c}_\pi^\dagger$ do not appear in the Hamiltonian.) 

After performing the last transformation, the quadratic pieces take the diagonal form 
\begin{equation} \label{def_Hplus}
\hat{H}^\pm_{\mathrm{XY}}=\sum_{\kappa\in\mathcal{K}^\pm}\varepsilon_\kappa\ \left(\hat{\eta}_\kappa^\dagger\hat{\eta}_\kappa-\frac{1}{2}\right)\,.
\end{equation}
We use the eigenvalues of the quasiparticle number operators $\hat{n}_\kappa = \hat{\eta}_\kappa^\dagger \hat{\eta}_\kappa^{}$ to write an orthonormal basis of eigenstates of $\hat{H}_{\mathrm{XY}}$. The eigenstates $\ket{\{n_\kappa\}}$ satisfy
\begin{equation}
\hat{n}_\kappa\ket{\{n_\kappa\}}=n_\kappa\ket{\{n_\kappa\}}
\end{equation}
with $n_\kappa\in \{0,1\}$. Here, $\kappa\in\mathcal{K}^\pm$ depending on whether the sum $\sum_{\kappa\in\mathcal{K}^\pm}n_\kappa$ is even ($+$) or odd ($-$). For later use, we also define the variables
\begin{equation}
N_\kappa=(-1)^{n_\kappa}\,,
\end{equation}
which provide an equivalent characterization of eigenstates as the variables $n_\kappa$. We emphasize that the quantity $N_\kappa$ is not related to the total number operator $\hat{N}=\sum^L_{j =1}\hat{f}_j ^\dagger\hat{f}_j $, which counts the total number of fermions.

The previous discussion shows that Hamiltonian~\eqref{eq:HXY1}, when written in terms of fermionic operators $\hat{f}_j$ and $\hat{f}^\dagger_j$, becomes two quadratic fermionic Hamiltonians over direct addends of the fermionic Hilbert space in Eq.~(\ref{eq:HXYn}). Therefore, the overall Hamiltonian is almost quadratic, but not quite. One has two distinct notions of quasiparticles, described by $\hat{H}_{\mathrm{XY}}^+$ and $\hat{H}_{\mathrm{XY}}^-$, depending on the number of quasiparticles. Namely, for an even number of quasiparticles one needs to use $\hat{H}_{\mathrm{XY}}^+$ while for an odd number of quasiparticles one needs to use $\hat{H}_{\mathrm{XY}}^-$. In particular, the ground state, which has no quasiparticles, is given by the ground state of $\hat{H}_{\mathrm{XY}}^+$, while the ground state of $\hat{H}_{\mathrm{XY}}^-$ is not an energy eigenstate of the system.

\section{Different averages over eigenstates}\label{app:evenodd}

Here we study how the eigenstates of Hamiltonians $\hat H^\pm$ in Eq.~(\ref{def_Hplus}), which are not eigenstates of Hamiltonian~(\ref{eq:HP}), change the average entanglement entropy. We define
\begin{equation} \label{def_deltaspm}
 \delta s_\pm(f) = \frac{\langle S_A \rangle_\pm - \langle S_A \rangle^\mathrm{all}_\pm}{L_A \ln 2} \,,
\end{equation}
where $\langle S_A \rangle_\pm$ are the partial averages, as defined in Eq.~(\ref{def_avrpm}), over eigenstates of the XY Hamiltonian~(\ref{eq:HP}) with only an even number of quasiparticles ($+$) or only an odd number of quasiparticles ($-$), respectively. On the other hand, $\langle S_A \rangle^\mathrm{all}_\pm$ are the averages over all eigenstates of Hamiltonians $\hat H^\pm$ in Eq.~(\ref{def_Hplus}). The Hamiltonian $\hat{H}_\mathrm{XY}$ is not quadratic, but it can be written as the sum of two quadratic Hamiltonians $\hat{H}_\mathrm{XY}^\pm$ projected onto the even/odd sectors $\mathcal{H}^\pm$ of the Hilbert space. We are therefore exploring how the average eigenstate entanglement entropy of $\hat{H}_{\mathrm{XY}}$ compares to computing first the average entanglement entropy for $\hat{H}^\pm$ individually over all their eigenstates (rather than just over the even or odd sectors) and then averaging the two.

\begin{figure}[!b]
\centering
\noindent\makebox[\linewidth]{
\includegraphics[width=0.98\linewidth]{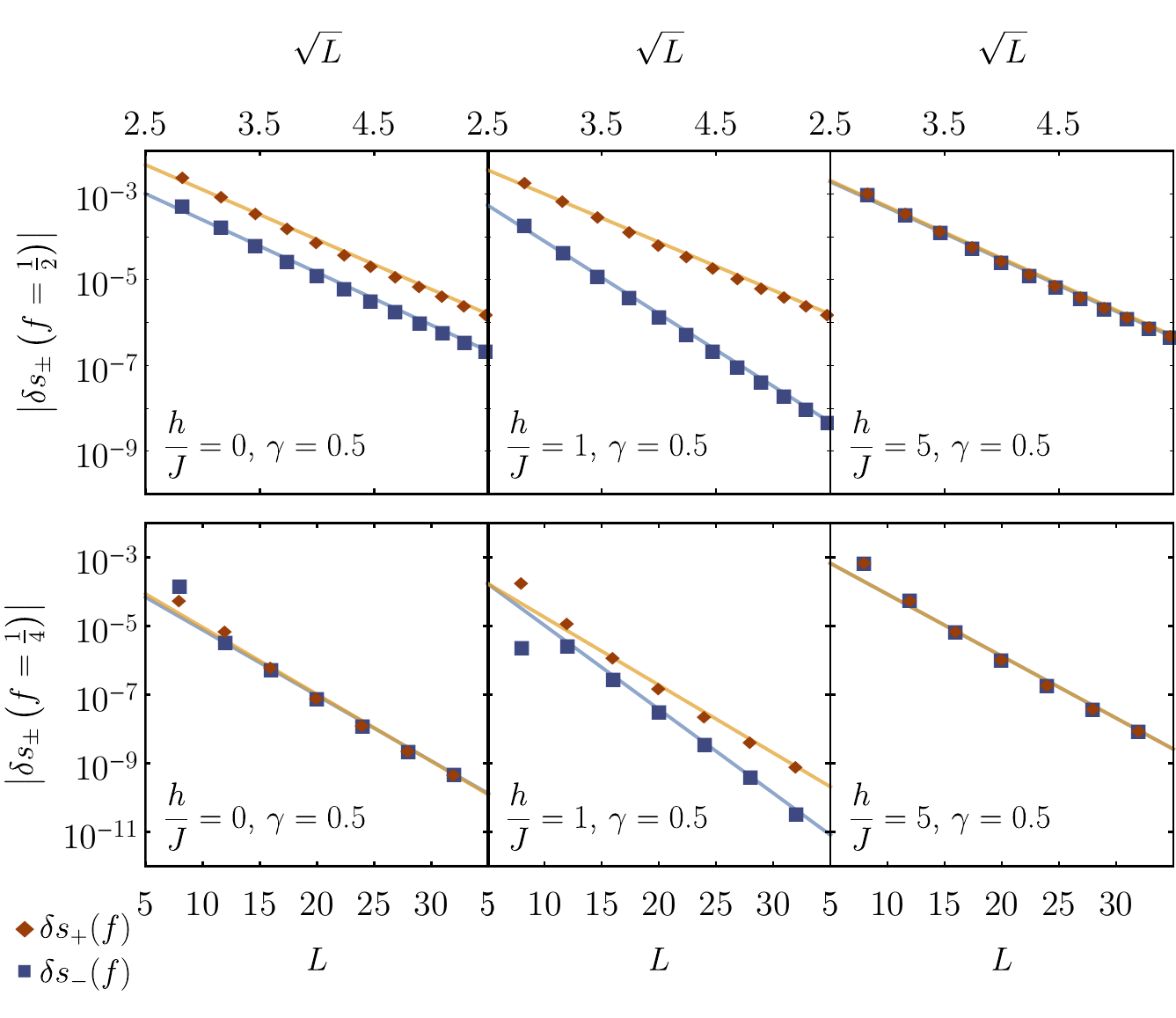}}
\vspace{-0.5cm}
\ccaption{Effect of averaging over ``non-physical'' eigenstates}{Differences $\delta s_\pm(f)$, see Eq.~(\ref{def_deltaspm}), vs $\sqrt{L}$ (top panels, for $f=1/2$) and vs $L$ (bottom panels, for $f=1/4$), for different values of $h/J$ and $\gamma=1/2$. The straight lines are fits that show that the differences scale $\propto\exp(-\alpha \sqrt{L})$ for $f=1/2$ and $\propto\exp(-\alpha' L)$ for $f=1/4$.}
\label{fig:pcbabc}
\end{figure}

Results for $|\delta s_\pm(f)|$ are shown in Fig.~\ref{fig:pcbabc} for $f=1/2$ (top panels) and $f=1/4$ bottom panels for different values of $h/J$ and $\gamma=1/2$. Those results reveal that the differences between the correct and the incorrect calculations vanish in the thermodynamic limit. We can prove this observation analytically by studying which terms in our series expansion in Eq.~(\ref{def_Sseries}) differ between the partial and the full average. 

Let us consider a binomial distribution of $L$ independent random variables $(N_1,\cdots,N_L)$ that can take the values $\pm 1$ with equal probability, i.e., a Rademacher distribution. For simplicity, we still assume $L$ to be an even number. The $n$-point correlation function $\langle N_{i_1}\cdots N_{i_n}\rangle$ is the one described in Eq.~(\ref{eq:binomial}): it either vanishes, or it is $1$ iff each variable $N_i$ appears an even number of times. If we now restrict to the partial averages $\langle \dots\rangle_\pm$ over all possible values $N_i$ subject to the constraint $\prod^L_{i=1}N_i=\pm 1$ ($L$ is still even), we find that the $n$-point correlation function only changes in the very special case in which each random variable $N_i$ appears an odd number of times, in which case we find
\begin{equation}
 \langle N_1\dots N_L\rangle_\pm=\pm 1,
\end{equation}
rather than $\langle N_1\dots N_L\rangle=0$. If we consider the series expansion of the entanglement entropy $S_A$ from Eq.~(\ref{def_Sseries}), the difference between the two averages is nonzero only for $n\geq L/2$, i.e., we find
\begin{equation}
 \delta s_\pm(f)=\frac{1}{L_A\ln{2}}\sum^\infty_{n=L/2}\frac{\langle\mathrm{Tr}\,[\ii \mathbb{J}]_A^{2n}\rangle_\pm-\langle\mathrm{Tr}\,[\ii \mathbb{J}]_A^{2n}\rangle_\pm^\mathrm{all}}{4n(2n-1)}\,.
\end{equation}
This implies that $\lim_{L\to\infty}\delta s_\pm(f)=0$, because the different averages agree to all order $n<L/2$. With this, we have proved that the differences vanish in the thermodynamic limit, while for the exact exponential scaling we rely on the numerical results shown in Fig.~\ref{fig:pcbabc}.

\section{Computation of higher order bounds for noninteracting fermions}\label{app1}

Here, we explain the method used to compute higher order bounds for noninteracting fermions [see Eq.~(\ref{def_Hfree})]. We also discuss the calculations of up to the fourth order bounds. They extend the results reported in Ref.~\cite{vidmar_hackl_17}, where we computed the first and second order bounds.

\subsection{Generalized Wick contractions and second order terms}\label{sec_traces}

In order to compute higher-order traces
\begin{equation}
 \langle\mathrm{Tr} [\ii \mathbb{J}]_A^{2n} \rangle=2\!\!\!\!\!\!\!\sum^{L_A}_{x_1,\cdots,x_{2n}=1}\!\!\!\!\!\!\!\langle j(x_1-x_2)\cdots j(x_{2n}-x_1)\rangle\,,\label{eq:def_trace_general}
\end{equation}
with $j(d_i)= \frac{1}{L} \sum_{\kappa\in\mathcal{K}}e^{\ii  \kappa d_i} N_\kappa$, we develop a systematic method of computing higher order correlation functions $\langle j(d_1)\cdots j(d_{2n})\rangle$. Technically, these are Fourier transformed correlation functions of the binomial distribution. They can be computed by adopting a strategy analogous to the one used for computing correlation functions of Gaussian distributions:

(i) The building blocks of the correlation functions are the so-called $2n$-contractions given by
\begin{equation}
 \contraction{}{j}{(d_1)j(d_2)\cdots j(d_{2n-1})}{j} \contraction{j(d_1)}{j}{(d_2)\cdots }{j} j(d_1)j(d_2)\cdots j(d_{2n-1})j(d_{2n})=\mathfrak{c}_{n}\frac{\bar{\delta}(\sum^{2n}_{i=1}d_i)}{L^{2n-1}}\,,
\end{equation}
where $\bar{\delta}(D)=1$ if $D=0\,\Mod{L}$ and it is zero otherwise, i.e., $D = \sum^{2n}_{i=1}d_i$ is restricted to be an integer multiple of $L$. The prefactors $\mathfrak{c}_{n}$ can be computed systematically as $\mathfrak{c}_n=L^{2n-1}\langle j(1)^{2n-1}j(1-2n)\rangle$ by using the explicit form $j(d_i)=\frac{1}{L}\sum_{\kappa\in\mathcal{K}}e^{\ii \kappa d_i}N_\kappa$. Evaluating this formula for the first few cases, one realizes that the $\mathfrak{c}_{n}$'s are given by the tangent numbers with alternating signs, namely
\begin{align}
 \mathfrak{c}_n&=\frac{1}{L}\hspace{-2em}\sum^L_{\qquad k_1,\ldots,k_{2n}=1}\exp{\frac{2\pi \ii}{L}\left(\sum^{2n}_{i=1}k_i-2n\,k_{2n}\right)}\nonumber\\
 &=\frac{2^{2 n} (2^{2 n}-1)}{2 n}\,B_{2 n}\,,
\end{align}
where $B_{2n}$ refers to the $2n$-th Bernoulli number. Up to fourth order, we only need
\begin{equation}
 \mathfrak{c}_1=1,\quad \mathfrak{c}_2=-2,\quad \mathfrak{c}_3=16,\quad\text{and}\quad \mathfrak{c}_4=-272.
\end{equation}
	
(ii) Once the $2n$-contractions are known, one can compute a general correlation function as
\begin{align}
\begin{split}
 \langle j(d_1)\cdots j(d_{2n})\rangle&=\sum(\text{all possible contractions})\\
 & =\contraction{}{j}{}{j}jj\cdots\contraction{}{j}{}{j}jj+\cdots+\contraction{}{j}{j\cdots j}{j} \contraction{j}{j}{\cdots }{j} jj\cdots jj\,,
\end{split}
\end{align}
where each contraction consists of a product of different pairings, quadruplings, etc., of the $2n$ $j$'s. This is a generalization of Wick's theorem because not only do we have $2$-contractions, but also higher-order $2n$-contractions.

To compute higher order traces up to order $n$, we need to determine all prefactors up to $\mathfrak{c}_n$. We can investigate the scaling in powers of $f=\frac{L_A}{L}$, which appear in $\lim_{L\to\infty}\langle\mathrm{Tr} [\ii \mathbb{J}]_A^{2n} \rangle/L_A$. A general $2n$ correlation function is schematically given by
\begin{align}
 &\langle j(d_1)\cdots j(d_{2n})\rangle=\contraction{}{j}{}{j}jj\cdots\contraction{}{j}{}{j}jj+\cdots+\contraction{}{j}{j\cdots j}{j} \contraction{j}{j}{\cdots }{j} jj\cdots jj \label{eq:GenCor} \\
 & = \mathfrak{c}_1^n\frac{\bar{\delta}(d_1+d_2)\cdots\bar{\delta}(d_{2n_1}+d_{2n})}{L^n} + \cdots + \mathfrak{c}_n\frac{\bar{\delta}(\sum^{2n}_{i=1}d_i)}{L^{2n-1}}\,.\nonumber
\end{align}
To compute $\langle \mathrm{Tr} [\ii \mathbb{J}]_A^{2n} \rangle/L_A$, we sum over $x_i$ with $1\leq i\leq 2n$, with $\sum^{2n}_{i=1}d_i$ automatically ensured to vanish, which implies that there is one redundant delta in each addend of Eq.~(\ref{eq:GenCor}).

The $2n$-correlation functions consists of various products of contractions. We refer to such a type of product of contractions as $(2n_1,\cdots,2n_l)$-contraction if it consists of $l$ contractions of types determined by $2n_i$. Moreover, we refer to the full set or sum of $(2n_1,\cdots,2n_l)$-contractions as $\mathcal{C}_{(2n_1,\cdots,2n_l)}$. A specific contraction $\mathfrak{C}\in\mathcal{C}_{(2n_1,\cdots,2n_l)}$ gives $l$ delta functions $\bar{\delta}(D_i)$ with $D_i$ being specific sums of the $d_i$'s according to the chosen contraction, namely,
\begin{equation}
\mathfrak{C}=\prod^l_{i=1}\mathfrak{c}_{n_i}\frac{\bar{\delta}(D_i)}{L^{2n_i-1}}\,.
\end{equation}
Provided that $D_i\leq L$, we can ignore the $\Mod{L}$ condition in $\bar{\delta}$ and replace it with a regular $\delta$. This product of deltas appearing in the sum (\ref{eq:def_trace_general}) then gives
\begin{equation}
\sum^{L_A}_{x_1,\cdots,x_{2n}=1}\hspace{-5mm}\delta(D_1)\cdots\delta(D_l)={\cal O}(L_A^{2n-l+1})\,,
\end{equation}
where we use the fact that each of the $l$ deltas reduces the dimension of the sum by one, except for the one that is redundant because $\sum^{l}_{i=1}D_i=\sum{2n}_{i=1}d_i=0$. Therefore, the sum gives a polynomial of degree $2n-l+1$. Applying this result to all contractions, we find
\begin{equation}
\sum^{L_A}_{x_1,\cdots,x_{2n}}\langle j(d_1)\cdots j(d_{2n})\rangle\sim\frac{{\cal O}(L_A^{n+1})}{L^n}+\cdots+\frac{{\cal O}(L_A^{2n})}{L^{2n-1}}
\end{equation}
as $f\to 0$, where ${\cal O}(L_A^{n+1})$ refers to a polynomial of degree $n+1$. Note that we included the requirement that all $D_i<L$ by considering this expression in the limit $f\to 0$. Therefore, we conclude that
\begin{equation}
\lim_{L\to\infty} \frac{\langle\mathrm{Tr} [\ii \mathbb{J}]_A^{2n} \rangle}{L_A}\sim\mathcal{O}(f^{n})+\cdots+\mathcal{O}(f^{2n-1})\quad\text{as}\quad f\to 0\,,
\end{equation}
i.e., that it is a polynomial function containing only powers from $f^{n}$ up to $f^{2n-1}$. This result is instrumental in the analysis of the behavior of the entropy for vanishing subsystem fraction $f \to 0$. It implies that it is sufficient to compute the expansion through order $n$ to get the dominant terms through order $f^n$. However, note that in general we expect additional corrections due to the $\Mod{L}$ condition in $\bar{\delta}$ for $f>\frac{1}{n}$. Interestingly, these corrections are absent in the relevant interval $f\in[0,\frac{1}{2}]$ for $n\leq 3$, but for $n=4$ we will see that additional terms do appear for $f>\frac{1}{4}$.\\

\noindent\textbf{Second order trace.} To illustrate this method, let us apply it to the second order correction containing $\langle\mathrm{Tr} [\ii \mathbb{J}]_A^{4} \rangle$. We need to compute the $4$-point correlation function $\langle j(d_1)j(d_2)j(d_3)j(d_4)\rangle$. We find three $2$-contractions and one $4$-contraction given by
\begin{widetext}
\begin{align}
\langle j(d_1)j(d_2)j(d_3)j(d_4)\rangle&=\mathcal{C}_{(2,2)}+\mathcal{C}_{(4)}=\left(\contraction{}{j}{}{j}
jj\contraction{}{j}{}{j}
jj+\contraction[2ex]{}{j}{jj}{j}\contraction{j}{j}{}{j}
jjjj+\contraction[2ex]{}{j}{j}{j}\contraction{j}{j}{j}{j}
jjjj\right)+\contraction{}{j}{jj}{j}\contraction{j}{j}{}{j}
jjjj\\
&=\mathfrak{c}_1^2\left(\frac{\bar{\delta}(d_1+d_2)\bar{\delta}(d_3+d_4)}{L^2}+\frac{\bar{\delta}(d_1+d_4)\bar{\delta}(d_2+d_3)}{L^2}+\frac{\bar{\delta}(d_1+d_3)\bar{\delta}(d_2+d_4)}{L^2}\right)+\mathfrak{c}_2\frac{\bar{\delta}(\sum^4_{i=1}d_i)}{L^3}\,.\nonumber
\end{align}
\end{widetext}
Plugging this expression into the sum~(\ref{eq:def_trace_general}) for the trace gives
\begin{equation}
\langle \mathrm{Tr}[\ii \mathbb{J}]^{4}_{A}\rangle=\frac{16L_A^3+2L_A}{3L^2}-\frac{4L_A^4}{L^3},
\end{equation}
as already found in Ref.~\cite{vidmar_hackl_17}.\footnote{Note that there was a missing factor of $2$ in the supplemental material of Ref.~\cite{vidmar_hackl_17}.} It then follows that
\begin{equation}
t_2(f)=\lim_{L\to\infty}\frac{\langle \mathrm{Tr}[\ii \mathbb{J}]^{4}_{A}\rangle}{L_A}=\frac{16}{3}f^2-4f^3,\label{eq:Order2}
\end{equation}
where the first term comes from the sum over the $2$-contractions, while the second term comes from the $4$-contraction for which $\bar{\delta}(\sum^4_{i=1}d_i)=1$ due to $\sum^4_{i=1}d_i=(x_1-x_2)+\cdots+(x_4-x_1)=0$. Note that, in Eq.~(\ref{eq:Order2}), we assume $L_A\leq L/2$ to avoid the complication that $\bar{\delta}(d_{i_1}+d_{i_2})$ is also nonzero for $d_{i_1}+d_{i_2}=0\,\Mod{L}$.

\subsection{Wick diagrams and third order terms}\label{app:Wick-diagrams}

When computing higher order correlation functions, we need to sum a larger number of contractions and it is important to keep track of all cases. Contractions can be efficiently organized by taking symmetry into account. The evaluation of a contraction in which one shifts all $d_i$'s by an overall constant does not change the result. It is important to emphasize that the $d_i$'s should be thought of as arranged in a circle. Hence, representing the equivalence class of contractions under translations is best done on a circle. We refer to such a pictorial representation as a Wick diagram
\begin{align}
\vcenter{\hbox{
		\begin{tikzpicture}
		\pgfmathsetmacro{\r}{.8}%
		\pgfmathsetmacro{\s}{1.3}%
		\pgfmathsetmacro{\step}{40}%
		\coordinate (o) at (0,0);
		\draw (o) circle (\r);
		\foreach \x in {1,...,3}{
			\fill (90+\step-\x*\step:\r) circle[radius=1pt];
			\draw (90+\step-\x*\step:\s) node{$d_\x$};
		}
		\fill (90+\step:\r) circle[radius=1pt];
		\draw (90+\step:\s) node{$d_{2n}$};
		\foreach \x in {8,...,16}{
			\draw (90+\step-\x*\step/2:\s) node{$\cdot$};
		}
		\end{tikzpicture}
}}\,,
\end{align}
where we connect the $2n$ dots associated to the $d_i$'s accordingly to get a specific Wick diagram. Each diagram $\mathfrak{C}\in\mathcal{C}_{(n_1,\cdots,n_l)}$ comes with some multiplicity $\#\mathfrak{C}$ associated to translations that do not lead to the same diagram. We refer to the underlying product of Kronecker deltas in a given diagram $\mathfrak{C}$ as
\begin{equation}
\delta_\mathfrak{C}=\delta(D_1)\cdots\delta(D_l)\,,
\end{equation}
where we ignore any contributions from $\Mod{L}$ in $\bar{\delta}(D_i)$. In the thermodynamic limit, we can evaluate the sum over $\delta_\mathfrak{C}$ as an integral
\begin{align}
\lim_{L\to\infty}\sum^{L_A}_{x_1,\cdots,x_{2n}=1}\hspace{-1mm}\delta_\mathfrak{C}&=\underset{0\,\,}{\overset{\,\,L_A}{\int}}d^{2n}x\,\delta(D_1)\cdots \delta(D_{l-1})\\
&=L_A^{2n+1-l}\,\underset{0\,\,}{\overset{\,\,1}{\int}}d^{2n}x\,\delta(D_1)\cdots \delta(D_{l-1})\,,\nonumber
\end{align}
by formally changing from Kronecker deltas to delta distributions. Note that we removed one of the Kronecker deltas due to the redundancy mentioned earlier, to avoid having an ill defined $\delta(0)$. Moreover, we define the dimensionless quantity
\begin{equation}
\int\mathfrak{C}=\int_0^1d^{2n}x\,\delta_\mathfrak{C},
\end{equation}
in which we remove the dependence on $L_A$ by making the replacement $x_i\to x_i\,L_A$, which only affects the range of integration. As long as we can ignore the $\Mod{L}$ condition in $\bar{\delta}$, i.e., for $f\in[0,\frac{1}{n}]$, we have
\begin{equation}
\lim_{L\to\infty}\frac{1}{L_A}\sum^{L_A}_{x_1,\cdots,x_{2n}=1}\mathfrak{C}=\mathfrak{c}_{2n_1}\cdots \mathfrak{c}_{2n_l}\,f^{2n-l}\,\int\mathfrak{C}\,.
\end{equation}
Finally, the summation over all diagrams within a class $\mathcal{C}_{(n_1,\cdots,n_l)}$ is given by
\begin{align}
&\int\mathcal{C}_{(n_1,\cdots,n_l)}=\sum_{\mathfrak{C}\in\mathcal{C}_{(n_1,\cdots,n_l)}}\lim_{L\to\infty}\frac{1}{L_A}\sum^{L_A}_{x_1,\cdots,x_{2n}=1}\mathfrak{C}\nonumber \\
&=\mathfrak{c}_{2n_1}\cdots \mathfrak{c}_{2n_l}\,f^{2n-l}\sum_{\mathfrak{C}\in\mathcal{C}_{(n_1,\cdots,n_l)}}\#\mathfrak{C}\!\int\mathfrak{C}\,.
\end{align}
Representing contractions by Wick diagrams, and replacing sums by integrals, will prove to be very efficient to evaluate the third and fourth order traces in the thermodynamic limit. Let us first reconsider the simpler example of the second order trace.\\

\noindent\textbf{Second order trace.} When we computed the terms of order $2$, we had four generalized Wick contractions. These correspond to only three Wick diagrams, namely
\begin{equation}
\contt{a}{c}{b}{d}\,,\quad\contt{a}{b}{c}{d}\,,\quad\conf{a}{b}{c}{d}\,,
\end{equation}
because the second diagram has multiplicity $2$. We can compute their prefactors as
\begin{align}
\int \contt{a}{c}{b}{d}&=\underset{0\,\,}{\overset{\,\,1}{\int}}d^{4}x\,\delta(x_1-x_2+x_3-x_4)=\frac{2}{3}\,,\\
\int\contt{a}{b}{c}{d}&=\underset{0\,\,}{\overset{\,\,1}{\int}}d^{4}x\,\delta(x_1-x_3)=1\,,\\
\int\conf{a}{b}{c}{d}&=\underset{0\,\,}{\overset{\,\,1}{\int}}d^{4}x=1\
\end{align}
of which only the first diagram requires some work. An efficient way of evaluating the integrals involved is to introduce a dummy variable $y$, leading to
\begin{align}
\int \contt{a}{c}{b}{d}&=\underset{0\,\,}{\overset{\,\,2}{\int}}dy\,\left[I_2(y)\right]^2,\quad I_2(y)=\underset{0\,\,}{\overset{\,\,1}{\int}}d^{2}x\,\delta(x_1+x_2-y)\,,
\end{align}
where we defined the piecewise polynomial function $I_2(y)$ over the intervals $[0,1]$ and $[1,2]$. Distinguishing between the two cases where $y$ is in either of the two intervals gives
\begin{equation}
I_2(y)=\left\{\begin{array}{ll}
I_2^{(1)}(y) & y\in[0,1]\\
I_2^{(1)}(2-y) & y\in[1,2]
\end{array}\right.\quad\text{with}\quad I_2^{(1)}(y)=\frac{y}{2}\,.
\end{equation}\\
Using this result, it is easy to compute $\frac{2}{3}$ as the prefactor. When adding up the contributions of the three diagrams, taking their multiplicity $\#\mathfrak{C}$, $\mathfrak{c}_1^2=1$, and the overall prefactor of $2$ into account, we find
\begin{equation}
t_2(f)=\lim_{L\to\infty}\frac{\langle\mathrm{Tr}[\ii \mathbb{J}]_A^{4}\rangle}{L_A}=\frac{16}{3}f^2-4f^3\,,
\end{equation}
which reproduces the result in Eq.~(\ref{eq:Order2}).\\

\noindent\textbf{Third order trace.} We can apply the representation of contractions as Wick diagrams to compute the third order terms. The main ingredients in the trace are
\begin{align}
\langle\mathrm{Tr}[\ii \mathbb{J}]_A^{6}\rangle&=2\sum^{L_A}_{x_1=1,\cdots,x_6=1}\langle j(d_1)\cdots j(d_6)\rangle\\
&=2\sum^{L_A}_{x_1=1,\cdots,x_6=1} \left[\mathcal{C}_{(2,2,2)} + \mathcal{C}_{(2,4)} + \mathcal{C}_{(6)}\right]\,.\nonumber
\end{align}

\begin{itemize}
\item \textbf{$(2,2,2)$-contraction: $\int\mathcal{C}_{(2,2,2)}=11\,\mathfrak{c}_1^3\,f^3=11\,f^3$}\\
The $15$ distinct contractions fall into $5$ different symmetry classes with multiplicities ranging from one to six. We list a representative $\mathfrak{C}$ of each class in table~\ref{tab:contract222}, and note its multiplicity $\#\mathfrak{C}$, and the prefactor $\int\mathfrak{C}$ resulting from the integration over the subsystem.
\item \textbf{$(2,4)$-contraction: $\int\mathcal{C}_{(2,4)}=12\,\mathfrak{c}_1\mathfrak{c}_2\,f^4=-24\,f^4$}\\
Due to the redundancy in Kronecker deltas, we can ignore the delta from the $4$-contraction and only focus on the $2$-contraction. There are $15$ distinct $2$-contractions, of which $6$ contract adjacent points, while the remaining $9$ do not. The prefactor is $1$ for the adjacent ones and $\frac{2}{3}$ for the remaining ones. We thus find $\int\mathcal{C}_{(2,4)}=(6\cdot 1+9\cdot\frac{2}{3})\,\mathfrak{c}_1\mathfrak{c}_2\,f^4=-24f^4$.
\item \textbf{$(6)$-contraction: $\int\mathcal{C}_{(6)}=\mathfrak{c}_3\,f^5=16\,f^5$}\\
This integration is trivial because the only contraction leads to a redundant delta leading to the prefactor $1$.
\end{itemize}

\begin{table}[!t]
\ccaption{$(2,2,2)$-contractions}{List of all $(2,2,2)$-contractions $\mathfrak{C}$, their multiplicity $\#\mathfrak{C}$ under translational symmetry, and their integrated prefactor $\int\mathfrak{C}$. Summing all terms, weighted with their multiplicity, gives $\int\mathcal{C}_{(2,2,2)}=11\,\mathfrak{c}_1^3\,f^3=f^3$.}
	\begin{center}
		\newcolumntype{C}{ >{\centering\arraybackslash} m{1cm} }
		\newcolumntype{B}{ >{\centering\arraybackslash} m{10cm} }
		\begin{tabular}{C|C|C||C|C|C}
			$\mathfrak{C}$ & $\#\mathfrak{C}$ & $\int\mathfrak{C}$ & $\mathfrak{C}$ & $\#\mathfrak{C}$ & $\int\mathfrak{C}$\\
			\hline\hline
			$\conttt{a}{b}{c}{d}{e}{f}$ & $2$ & $1$ & $\conttt{a}{c}{b}{e}{d}{f}$ & $3$ & $\frac{1}{2}$\\
			$\conttt{a}{b}{c}{f}{e}{d}$ & $3$ & $1$ & $\conttt{a}{d}{b}{e}{c}{f}$ & $1$ & $\frac{1}{2}$\\
			$\conttt{a}{b}{c}{e}{d}{f}$ & $6$ & $\frac{2}{3}$ & & &
		\end{tabular}
	\end{center}
\label{tab:contract222}
\end{table}

Combining all the terms computed above, and including the prefactor of $2$, we find
\begin{equation}
t_3(f)=\lim_{L\to\infty}\frac{\langle\mathrm{Tr}[\ii \mathbb{J}]^6_A\rangle}{L_A}=22 f^3-48f^4+32f^5\,.
\end{equation}

\subsection{Delta terms, fourth order terms, and\\ non-analytic contributions}\label{ssec:fourth-order}

An important subtlety arises starting with the fourth order bound, because it matters that our generalized Kronecker deltas are defined $\Mod L$. This means that we have $\delta(D)=1$ not only for $D=0$, but also for any multiple of the full system size $L$. We will see that this leads to surprising effects and turns the bounds into piecewise defined polynomials rather than a single analytic polynomial for $f\in[0,\tfrac{1}{2}]$. Such a correction is expected to appear for $f>\frac{1}{n}$ for traces $t_n$ of order $n$. This subtlety comes from the fact that our generalized Kronecker deltas are defined as
\begin{equation}
\bar{\delta}(D)=\left\{\begin{array}{cl}
1 & D=0\,\Mod{L}\\
0 & \text{else}
\end{array}\right.\,,
\end{equation}
where $D=\sum^{2j}_{i=1}d_{l_i}$ with $d_l=x_l-x_{l+1}$ and $x_{L+1}=x_1$. Up to third order, the $\Mod L$ condition played no role. The explanation for this consists of two parts:

(i) For a $2$-contraction $\contraction{}{j}{(d_{l})}{j}j(d_l)j(d_{m})=\frac{\mathfrak{c}_1}{L}\delta(x_l-x_{l+1}+x_m-x_{m+1})$, the largest possible value of $D=d_l+d_m$ is given by $D_{\max}=2(L_A-1)$, namely for $x_l=x_m=L_A$ and $x_{l+1}=x_{m+1}=1$. However, we only compute the trace for $L_A\leq L/2$ leading to $D_{\max}\leq L-2<L$. Thus for any $2$-contraction, we can just ignore the $\Mod L$ condition altogether.

(ii) Based on the first statement, we only need to take the $\Mod L$ condition into account for $\delta(\sum^{2j}_{i=1}d_{l_i})$ when $j>2$. Such generalized Kronecker deltas only appear for $4$-contractions and higher orders. Moreover, we should recall that at bound order $n$, we always have $\sum^{2n}_{i=1}d_i=0$, which implies that we can always choose one generalized Kronecker delta to be redundant. In the second and third order bounds, we do have terms that contain $4$- and $6$-contractions, but they always appear alone or in a product with only $2$-contractions. In those cases, we can always choose the $4$- or $6$-contraction to be the redundant one and we are left with $2$-contractions, discussed in point (i) above.

To understand how to take the $\Mod L$ condition into account, as well as its effects, let us consider the fourth order trace.\\

\noindent\textbf{Fourth order trace.} The fourth order bound is computed from
\begin{equation}
\langle\mathrm{Tr}[\ii \mathbb{J}]_A^{8}\rangle=2\sum^{L_A}_{x_1=1,\cdots,x_8=1}\langle j(d_1)\cdots j(d_8)\rangle
\end{equation}
Using again the generalized Wick's theorem, the structure of the $8$-point correlation function is given by
\begin{align}
\langle j(d_1)\cdots j(d_8)\rangle = &\mathcal{C}_{(2,2,2,2)} + \mathcal{C}_{(2,2,4)} \\&+\left(\mathcal{C}_{(4,4)}+\mathcal{C}_{(2,6)}\right)+\mathcal{C}_{(8)}\,,\nonumber
\end{align}
where we grouped the $(4,4)$-contractions together with the $(2,6)$-contractions, because they contribute a power $L_A^7$. Using straightforward combinatorics, one can count how many contractions there are per type,
\begin{align}
\begin{split}
\#\mathcal{C}_{(2,2,2,2)}=105\,,\quad\#\mathcal{C}_{(2,2,4)}=210\,,\hspace{.5cm}\\
\#\mathcal{C}_{(4,4)}=70\,,\quad\#\mathcal{C}_{(2,6)}=28\,,\quad\#\mathcal{C}_{(8)}=1\,,
\end{split}
\end{align}
leading to a total of $414$ distinct contractions. As before, many contractions are related by translational symmetry, so it is sufficient to compute one representative per symmetry class.

\begin{table}[!t]
\ccaption{$(2,2,2,2)$-contractions}{List of all $(2,2,2,2)$-contractions $\mathfrak{C}$, their multiplicity  $\#\mathfrak{C}$ under translational symmetry, and their prefactor $\int\mathfrak{C}$ under integration. Summing all terms, weighted with their multiplicity, gives $\int\mathcal{C}_{(2,2,2,2)}=\tfrac{908}{15}\,\mathfrak{c}_1^4\,f^4=\tfrac{908}{15}\,f^4$.}
	\begin{center}
		\newcolumntype{C}{ >{\centering\arraybackslash} m{1.2cm} }
		\begin{tabular}{C|C|C||C|C|C}
			$\mathfrak{C}$ & $\#\mathfrak{C}$ & $\int\mathfrak{C}$ & $\mathfrak{C}$ & $\#\mathfrak{C}$ & $\int\mathfrak{C}$\\
			\hline\hline
			$\contttt{a}{d}{b}{c}{e}{f}{g}{h}$		& $8$ 	& $1$				&	%
			$\contttt{a}{d}{c}{b}{e}{h}{g}{f}$		& $4$ 	& $1$				\\	%
			$\contttt{a}{b}{c}{d}{e}{f}{g}{h}$		& $2$	& $1$				&	%
			$\contttt{a}{c}{b}{d}{e}{f}{g}{h}$ 	& $8$	& $\tfrac{2}{3}$	\\
			$\contttt{a}{c}{b}{d}{e}{g}{f}{h}$ 	& $8$	& $\tfrac{2}{3}$	&	%
			$\contttt{a}{c}{b}{f}{d}{e}{g}{h}$ 	& $8$	& $\tfrac{2}{3}$	\\	%
			$\contttt{a}{e}{b}{f}{c}{d}{g}{h}$ 		& $4$	& $\tfrac{2}{3}$	&	%
			$\contttt{a}{c}{b}{e}{d}{g}{f}{h}$ 		& $8$	& $\tfrac{11}{30}$	\\
			$\contttt{a}{c}{b}{f}{d}{g}{e}{h}$ 		& $8$	& $\tfrac{11}{30}$	&
			$\contttt{b}{f}{c}{g}{e}{h}{a}{d}$	& $4$ 	& $\tfrac{11}{30}$	\\
			$\contttt{a}{d}{b}{g}{e}{h}{c}{f}$ 	& $2$	& $\tfrac{11}{30}$	&	%
			$\contttt{a}{c}{b}{e}{d}{f}{g}{h}$ 	& $8$	& $\tfrac{1}{2}$	\\	%
			$\contttt{a}{c}{b}{e}{d}{h}{g}{f}$		& $8$ 	& $\tfrac{1}{2}$	&	%
			$\contttt{a}{c}{b}{g}{d}{h}{f}{e}$ 		& $8$	& $\tfrac{1}{2}$	\\
			$\contttt{a}{d}{b}{e}{c}{f}{g}{h}$ 		& $8$	& $\tfrac{1}{2}$	&	%
			$\contttt{a}{c}{b}{d}{e}{g}{f}{h}$ 	& $4$	& $\tfrac{9}{20}$	\\	%
			$\contttt{a}{c}{b}{f}{d}{h}{g}{e}$	& $4$ 	& $\tfrac{2}{5}$	&	%
			$\contttt{a}{e}{b}{f}{c}{g}{d}{h}$ 	& $1$	& $\tfrac{2}{5}$		%
		\end{tabular}
	\end{center}
\label{tab:contract2222}
\end{table}

\begin{itemize}
\item \textbf{$(2,2,2,2)$-contractions: $\int \mathcal{C}_{(2,2,2,2)}=\frac{908}{15}\,f^4$}\\
The $105$ distinct contractions fall into $18$ different symmetry classes, with multiplicities ranging from one to eight. We list a representative $\mathfrak{C}$ of each class in table~\ref{tab:contract2222} and note both, its multiplicity $\#\mathfrak{C}$ and its prefactor $\int\mathfrak{C}$ under integration over the subsystem.
\item \textbf{$(2,2,4)$-contractions: $\int \mathcal{C}_{(2,2,4)}=-\frac{1296}{5}\,f^5$}\\
The $210$ distinct contractions fall into $29$ different symmetry classes, with multiplicities ranging from one to eight. We list a representative $\mathfrak{C}$ of each class in table~\ref{tab:contract224} and note both, its multiplicity $\#\mathfrak{C}$ and its prefactor $\int\mathfrak{C}$ under integration over the subsystem.
\item \textbf{$(4,4)$-contractions: $\int\mathcal{C}_{(4,4)}=\frac{5816}{63}\,f^6$}\\
The $70$ distinct contractions fall into $7$ different symmetry classes, with multiplicities ranging from one to eight. We list a representative $\mathfrak{C}$ of each class in table~\ref{tab:contract44} and note both, its multiplicity $\#\mathfrak{C}$ and its prefactor $\int\mathfrak{C}$ under integration over the subsystem.
\item \textbf{$(2,6)$-contractions: $\int \mathcal{C}_{(2,6)}=\frac{1024}{3}\,f^6$}\\
Similar to $\mathcal{C}_{(2,4)}$, we can ignore the delta from the $6$-contraction and only focus on the $2$-contraction. There are $8\cdot 7/2=28$ distinct $2$-contractions, of which $8$ contract adjacent points, while the remaining $20$ do not. As before, the prefactor for adjacent contractions is $1$, and it is $\frac{2}{3}$ for the remaining ones. Overall, we thus find $\int\mathcal{C}_{(2,4)}=(8\cdot 1+20\cdot\frac{2}{3})\,\mathfrak{c}_1\mathfrak{c}_3\,f^6=\frac{1024}{3}\,f^6$.
\item \textbf{$(8)$-contractions: $\mathfrak{c}_4=\int\mathcal{C}_{(8)}=\mathfrak{c}_4=-272\,f^7$}\\
This integration is trivial because the only contraction gives a redundant delta leading to the prefactor $1$.
\end{itemize}

\begin{table}[!t]
\ccaption{$(2,2,4)$-contractions}{List of all $(2,2,4)$-contractions $\mathfrak{C}$, their multiplicity $\#\mathfrak{C}$ under translational symmetry, and their prefactor $\int\mathfrak{C}$ under integration. Summing all terms, weighted with their multiplicity, gives $\mathcal{C}_{(2,2,4)}=\frac{648}{5}\,\mathfrak{c}_1^2\mathfrak{c}_2\,f^5=-\frac{1296}{5}\,f^5$.}
\vspace{-0.3cm}
	\begin{center}
		\newcolumntype{C}{ >{\centering\arraybackslash} m{1.1cm} }
		\newcolumntype{K}{ >{\centering\arraybackslash} m{6cm} }
		\begin{tabular}{K|C|C}
			$\mathfrak{C}$ & $\#\mathfrak{C}$ & $\int\mathfrak{C}$ \\
			\hline\hline
			$\conttf{a}{b}{c}{d}{e}{f}{g}{h},\conttf{a}{d}{b}{c}{e}{f}{g}{h},\conttf{a}{b}{d}{e}{c}{f}{g}{h}$ & $8$ & $1$ \\
			$\conttf{a}{b}{e}{f}{c}{d}{g}{h}$ & $4$ & $1$ \\
			$\conttf{a}{c}{b}{d}{e}{f}{g}{h},\conttf{a}{c}{d}{e}{b}{f}{g}{h},\conttf{a}{c}{g}{h}{f}{b}{d}{e},\conttf{a}{c}{e}{f}{b}{d}{g}{h},\conttf{a}{b}{d}{f}{c}{e}{g}{h}$ & $8$ & $\frac{2}{3}$ \\
			$\conttf{b}{e}{h}{g}{a}{c}{d}{f},\conttf{a}{e}{c}{d}{b}{f}{g}{h},\conttf{g}{c}{a}{h}{f}{b}{d}{e},\conttf{c}{f}{g}{h}{a}{b}{d}{e},\conttf{c}{h}{g}{f}{a}{b}{d}{e}$ & $8$& $\frac{2}{3}$ \\
			$\conttf{b}{g}{c}{h}{a}{d}{e}{f},\conttf{b}{h}{c}{g}{a}{d}{e}{f},\conttf{b}{d}{h}{c}{a}{e}{f}{g},\conttf{b}{g}{h}{f}{a}{c}{d}{e},\conttf{c}{g}{h}{f}{a}{b}{d}{e}$ & $8$ & $\frac{1}{2}$ \\
			$\conttf{a}{e}{b}{f}{c}{d}{g}{h},\conttf{a}{f}{e}{b}{c}{d}{g}{h}$ & $4$ & $\frac{1}{2}$ \\
			$\conttf{c}{g}{e}{h}{a}{b}{d}{f},\conttf{c}{f}{d}{h}{a}{b}{e}{g},\conttf{c}{h}{e}{g}{a}{b}{d}{f},\conttf{c}{h}{f}{d}{a}{b}{e}{g},\conttf{b}{g}{e}{h}{a}{c}{d}{f},\conttf{b}{h}{e}{g}{a}{c}{d}{f}$ & $8$ & $\frac{9}{20}$ \\
			$\conttf{a}{c}{e}{g}{b}{d}{f}{h}$ & $4$ & $\frac{4}{9}$ \\
			$\conttf{a}{e}{c}{g}{b}{d}{f}{h}$ & $2$ & $\frac{4}{9}$	
		\end{tabular}
	\end{center}
\label{tab:contract224} 
\end{table}

\begin{table}[!b]
\ccaption{$(4,4)$-contractions}{List of all $(4,4)$-contractions $\mathfrak{C}$, their multiplicity $\#\mathfrak{C}$ under translational symmetry, and their prefactor $\int\mathfrak{C}$ under integration. Summing all terms, weighted with their multiplicity, gives $\mathcal{C}_{(4,4)}=\tfrac{1454}{63}\,\mathfrak{c}_2^2\,f^6=\frac{5816}{63}\,f^6$.}
	\begin{center}
		\newcolumntype{C}{ >{\centering\arraybackslash} m{1.1cm} }
		\newcolumntype{K}{ >{\centering\arraybackslash} m{6cm} }
		\begin{tabular}{C|C|C||C|C|C}
			$\mathfrak{C}$ & $\#\mathfrak{C}$ & $\int\mathfrak{C}$ & $\mathfrak{C}$ & $\#\mathfrak{C}$ & $\int\mathfrak{C}$ \\
			\hline\hline
			$\conff{a}{b}{c}{d}{e}{f}{g}{h}$		& $4$ 	& $1$	&
			$\conff{a}{c}{d}{e}{b}{f}{g}{h}$		& $8$ 	& $\tfrac{2}{3}$ \\
			$\conff{a}{b}{d}{e}{c}{f}{g}{h}$		& $8$ 	& $\tfrac{2}{3}$	&
			$\conff{a}{b}{e}{f}{c}{d}{g}{h}$		& $2$ 	& $\tfrac{2}{3}$ \\
			$\conff{a}{c}{e}{f}{b}{d}{g}{h}$		& $8$ 	& $\tfrac{11}{20}$	&
			$\conff{a}{c}{d}{f}{b}{e}{g}{h}$ & $4$ 	& $\tfrac{11}{20}$	\\
			$\conff{a}{c}{e}{g}{b}{d}{f}{h}$ & $1$ & $\tfrac{151}{315}$
		\end{tabular}
	\end{center}
\label{tab:contract44}
\end{table}

\begin{figure*}[!t]
	\noindent
	\begin{minipage}{.45\textwidth}
		\includegraphics[width=\textwidth]{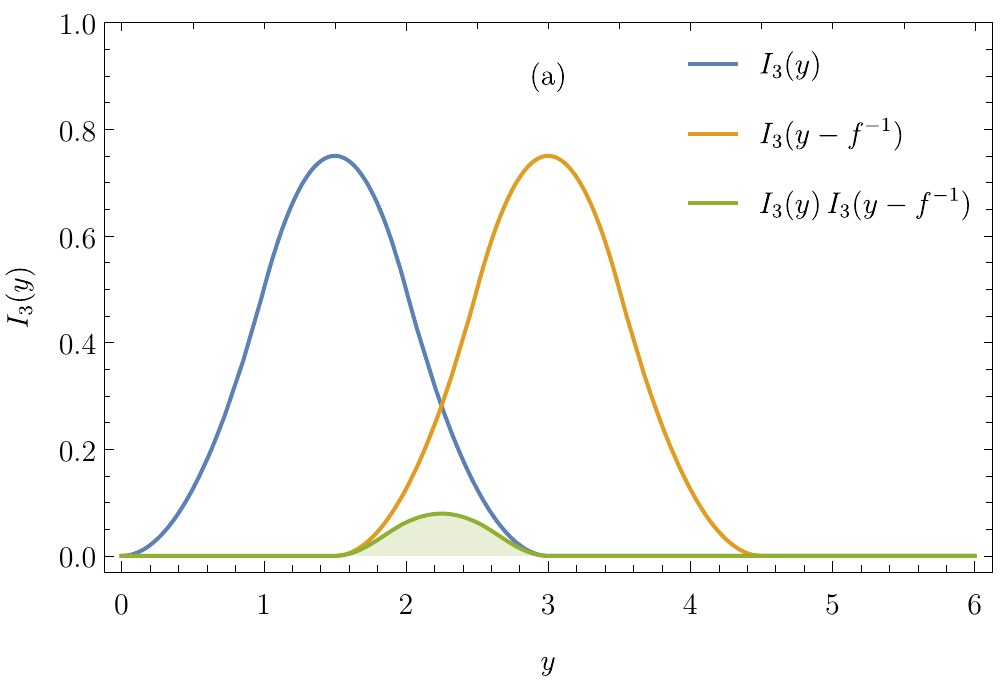}
	\end{minipage}
	\hspace{.04\textwidth}
	\begin{minipage}{.45\textwidth}
		\includegraphics[width=\textwidth]{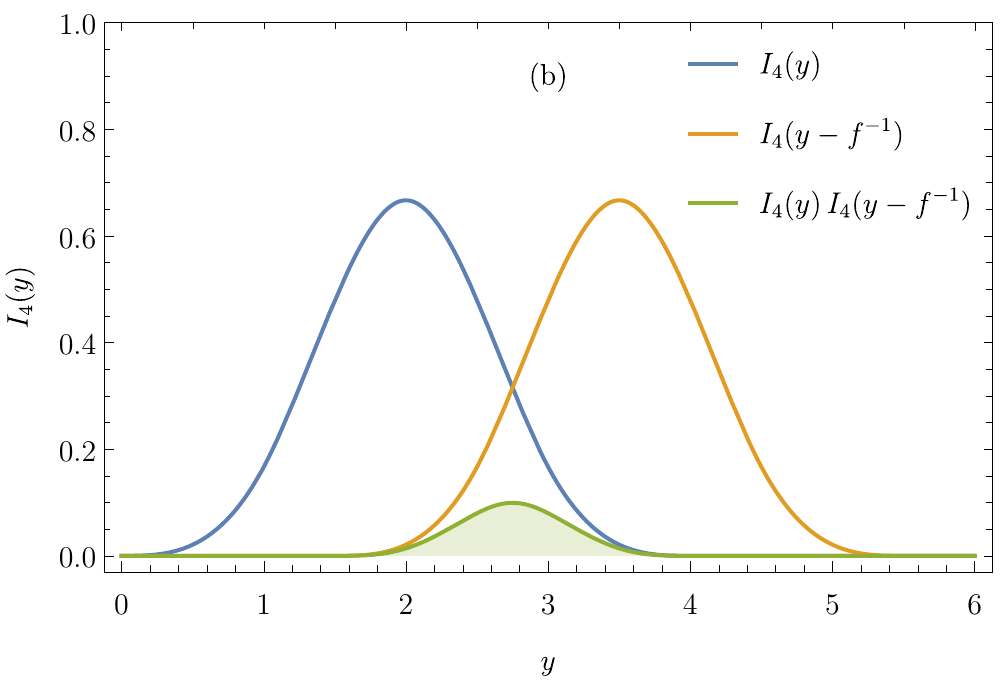}
	\end{minipage}
	\vspace{-0.2cm}
	\ccaption{Integrand $I_3(y)$ and $I_4(y)$}{This figure shows the integrands (a) $I_3(y)$ and (b) $I_4(y)$. When integrating their product $I_n(y)I_n(y-f^{-1})$ with a relative shift $f^{-1}$, we compute the area of the shaded region. For the illustration, we chose $f=2/3$.}
	\label{fig:integrals-I4-I3}
\end{figure*}

Putting all terms together results in the fourth order trace
\begin{equation}\label{eq:t4app}
t_4(f)=\frac{1816}{15}\,f^4-\frac{2592}{5}f^5+\frac{54640}{63}f^6-544 f^7\,.
\end{equation}

However, Eq.~\eqref{eq:t4app} only holds for $r\leq \frac{1}{4}$. The reason for this is exactly the subtlety mentioned before, namely, there are diagrams for which $\bar{\delta}(D_i)$ actually gives contributions due to the $\Mod L$ condition. Clearly, only $(4,4)$-contractions can give such a contribution, because all other contractions only consist of $2$-contractions plus $4$- or $6$-contractions that we can choose to be the redundant ones. Going through the list of $(4,4)$-contractions in table~\ref{tab:contract44}, we find the following contractions that lead to $\bar{\delta}(D_i)$ with $D_i=\pm L$ for $f<\frac{1}{2}$:
\begin{align}
\conff{a}{c}{e}{g}{b}{d}{f}{h}&=\frac{\bar{\delta}(x_1-x_2+x_3-x_4+x_5-x_6+x_7-x_8)}{L^6}\,,\label{eq:sub-case4}\\
\conff{a}{c}{e}{f}{b}{d}{g}{h}&=\frac{\bar{\delta}(x_1-x_2+x_3-x_4+x_5-x_7)}{L^6}\label{eq:sub-case2}\,,\\
\conff{a}{c}{d}{f}{b}{e}{g}{h}&=\frac{\bar{\delta}(x_1-x_2+x_3-x_5+x_6-x_7)}{L^6}\label{eq:sub-case3}\,.
\end{align}
With our computation of $\int\mathfrak{C}$, as listed in table~\ref{tab:contract44}, we already covered the case where $D_i=0$. However, for the Wick diagrams in Eq.~(\ref{eq:sub-case4}), we also need to compute the contributions from
\begin{equation}\label{eq:i4app}
\int^{L_A}_0d^8x\delta(d_1+d_3+d_5+d_7\pm L)=\int_0^4 dy\,I_4(y)I_4(y\pm f^{-1})\,,
\end{equation}
where $I_4(y)=\int_0^1d^4x\,\delta(x_1+x_2+x_3+x_4-y)$. Here, we use again the trick of introducing a dummy variable $y$ and rescaling by $L_A$, such that the $\Mod L$ condition becomes $\mod{\frac{L}{L_A}}=\mod{f^{-1}}$. Computing the function $I_4(f)$ requires some work because one needs to differentiate between several cases. Ultimately, we find a piecewise defined polynomial, with compact support on the interval $[0,4]$, given by
\begin{equation}
I_4(y)=\left\{\begin{array}{ll}
I^{(1)}_4(y) & y\in[0,1]\\
I^{(2)}_4(y) & y\in[1,2]\\
I^{(2)}_4(4-y) & y\in[2,3]\\
I^{(1)}_4(4-y) & y\in[3,4]
\end{array}\right.,
\end{equation}
with $I^{(1)}_4(y)=\frac{y^3}{6}$, and $I^{(2)}(y)=\frac{1}{6}(-3 y^3+12 y^2-12 y+4)$. 

One can deal with the Wick diagrams in Eqs.~\eqref{eq:sub-case2} and~\eqref{eq:sub-case3} in an analogous way, we find
\begin{equation}\label{eq:i3app}
\int^{L_A}_0d^8x\delta(d_1+d_3+d_5+d_7\pm L)=\int_0^3 dy\,I_3(y)I_3(y\pm f^{-1})\,,
\end{equation}
where the integral of both diagrams can be identified by renaming the variables $x_i$. We introduced $I_3(y)=\int_0^1d^4x\,\delta(x_1+x_2+x_3-y)$, which is given by
\begin{equation}
I_3(y)=\left\{\begin{array}{ll}
I^{(1)}_3(y) & y\in[0,1]\\
I^{(2)}_3(y) & y\in[1,2]\\
I^{(1)}_3(3-y) & y\in[2,3]\\
\end{array}\right.,
\end{equation}
with $I^{(1)}_3(y)=\frac{y^2}{2}$ and $I^{(2)}_3(y)=-y^2+3y-\frac{3}{2}$. 

The integrands in Eqs.~\eqref{eq:i4app} and~\eqref{eq:i3app} are therefore given by the product of two piecewise defined polynomials, as illustrated in Fig.~\ref{fig:integrals-I4-I3}. The integrals can be computed analytically in each interval, and are given by the piecewise defined functions:
\begin{widetext}
\begin{align}
\int_0^3 dy\,I_3(y)I_3(y\pm f^{-1})&=\left\{\begin{array}{ll}
-\frac{1}{120 f^5}+\frac{1}{8 f^4}-\frac{3}{4 f^3}+\frac{9}{4 f^2}-\frac{27}{8 f}+\frac{81}{40} & f\in[\frac{1}{3},\frac{1}{2}]\\[2mm]
0 & \text{else}
\end{array}\right.\,,\\
\int_0^4 dy\,I_4(y)I_4(y\pm f^{-1})&=\left\{\begin{array}{ll}
-\frac{1}{5040 f^7}+\frac{1}{180 f^6}-\frac{1}{15 f^5}+\frac{4}{9 f^4}-\frac{16}{9 f^3}+\frac{64}{15 f^2}-\frac{256}{45 f}+\frac{1024}{315} & f\in[\frac{1}{4},\frac{1}{3}]\\[2mm]
\frac{1}{720 f^7}-\frac{1}{36 f^6}+\frac{7}{30 f^5}-\frac{19}{18 f^4}+\frac{49}{18 f^3}-\frac{23}{6 f^2}+\frac{217}{90 f}-\frac{139}{630} & f\in[\frac{1}{3},\frac{1}{2}]\\[2mm]
0 & \text{else}
\end{array}\right.\,.
\end{align}
\end{widetext}

For higher order $n$, the non-analytic part of $t_n$ will contain integrals of the form
\begin{align}\label{eq:In-overlap}
 \int_0^ndy\,I_n(y)I_n(y\pm m f^{-1})\quad\text{with}\quad m\in\mathbb{N}_{>0}\,.
\end{align}
The relevant functions $I_n$ will be compactly supported in the region $[0,n]$, such that the integral in Eq.~(\ref{eq:In-overlap}) will vanish for $m f^{-1}>n$, i.e., $f<m/n$. The smallest possible value is $m=1$. Thus, we only expect non-analytical terms in $t_n$ for $f>\frac{1}{n}$.

To compute the correction for $t_4(f)$, let us recall that each of these terms comes with an additional multiplicity of $2$ due to the two different choices $\pm L$ in the $\Mod{L}$ condition, which can be translated into two distinct choices of $\pm f^{-1}$ in our rescaled variables. Furthermore, we need to take the prefactors $\mathfrak{c}_2^2=4$ and the diagram multiplicities
\begin{equation}
\#\conff{a}{c}{e}{g}{b}{d}{f}{h}=1\,,\quad\#\conff{a}{c}{e}{f}{b}{d}{g}{h}=8\,,\quad\#\conff{a}{c}{d}{f}{b}{e}{g}{h}=4
\end{equation}
into account. With this in hand, we find the correction term $t^{\mathrm{cor}}_4(f)$ introduced in Eq.~(\ref{def_s4m}) of the main text:
\begin{widetext}
\begin{equation} \label{def_tcorr}
t^{\mathrm{cor}}_4(f)=\left\{\begin{array}{ll}
0 & f\in[0,\frac{1}{4}]\\[2mm]
\frac{16384 f^6}{315}-\frac{4096 f^5}{45}+\frac{1024 f^4}{15}-\frac{256 f^3}{9}+\frac{64 f^2}{9}-\frac{16 f}{15}+\frac{4}{45}-\frac{1}{315f} & f\in[\frac{1}{4},\frac{1}{3}]\\[2mm]
\frac{24272 f^6}{63}-\frac{27424 f^5}{45}+\frac{1112 f^4}{3}-\frac{904 f^3}{9}+\frac{64 f^2}{9}+\frac{32 f}{15}-\frac{4}{9}+\frac{1}{45f} \qquad \qquad & f\in[\frac{1}{3},\frac{1}{2}]\\[2mm]
\end{array}\right.
\end{equation}
\end{widetext}
Note that the trace formula~(\ref{eq:def_trace_general}) contains an additional factor of $2$, and we need to include again the power $f^7$, which we separated from the integral by rescaling $x_i\to x_i\,L_A$.

\bibliographystyle{biblev1}
\bibliography{references}

\end{document}